\documentclass[aps,prl,twocolumn,showpacs,superscriptaddress,groupedaddress,nofootinbib]{revtex4-2}  
\usepackage{cancel}
\usepackage{graphicx} 
\usepackage[T1]{fontenc}
\usepackage{lmodern} 
\usepackage{graphicx}  
\usepackage{dcolumn}   
\usepackage{bm}        
\usepackage{amssymb}   
\usepackage{amsmath}
\usepackage{dsfont}
\usepackage{bbold}
\usepackage{array}
\usepackage{float}
\usepackage{makecell}
\usepackage{braket}
\usepackage{longtable}
\usepackage{supertabular,booktabs}
\usepackage{soul}

\usepackage{titlesec} 
\usepackage[colorlinks,citecolor=blue,urlcolor=blue,hypertexnames=true]{hyperref}
\usepackage{subfigure}

\usepackage[usenames,dvipsnames,svgnames,table]{xcolor} 

\renewcommand{\arraystretch}{2}

\newcommand{\be}{\begin{equation}}
\newcommand{\ee}{\end{equation}}
\newcommand{\bea}{\begin{eqnarray}}
\newcommand{\eea}{\end{eqnarray}}
\newcommand{\bml}{\begin{subequations}}
\newcommand{\eml}{\end{subequations}}
\newcommand{\bfig}{\begin{figure}}
\newcommand{\efig}{\end{figure}}

\newcommand{\slashed}{\hspace{-1.1ex}/}

\newcommand{\bmat}{\begin{pmatrix}}
\newcommand{\emat}{\end{pmatrix}}
\usepackage{graphicx, slashed,booktabs, color, multirow, float,
amsfonts, bbold, mathtools, sidecap, tikz, bm,enumitem}
\usepackage{multirow}
\usepackage{bbding}
\usepackage{titlesec}
\usepackage{hyperref}
\usepackage{wasysym}
\usepackage{amssymb}
\usepackage{pifont}

\renewcommand{\leq}{\leqslant}
\renewcommand{\geq}{\geqslant}

\usepackage[dvipsnames, usenames]{xcolor}

\definecolor{linkcolor}{rgb}{0.55, 0.13, .32}

\definecolor{oucrimsonblack}{rgb}{0.6, 0.0, 0.0}
\definecolor{persianblue}{rgb}{0.11, 0.22, 0.73}
\definecolor{forestgreen}{rgb}{0.13,0.35,0.13}
\definecolor{lightgray}{rgb}{0.83, 0.83, 0.83}
 \hypersetup{colorlinks, citecolor=oucrimsonred, linkcolor=persianblue, urlcolor=oucrimsonred}
\definecolor{cornellred}{rgb}{0.7, 0.11, 0.11}
\definecolor{navyblue}{rgb}{0.0, 0.0, 0.5}
\definecolor{amethyst}{rgb}{0.6, 0.4, 0.8}
\definecolor{yellow}{rgb}{1.0, 1.0, 0.0}
\definecolor{firebrick}{rgb}{0.7, 0.13, 0.13}
\definecolor{tangerineyellow}{rgb}{1.0, 0.8, 0.0}
\definecolor{deepfuchsia}{rgb}{0.76, 0.33, 0.76}
\definecolor{amber}{rgb}{1.0, 0.75, 0.0}
\definecolor{VioletRed4}{rgb}{0.55, 0.13, .32}
\definecolor{indiagreen}{rgb}{0.07, 0.53, 0.03}
\definecolor{VioletRed4}{rgb}{0.55, 0.13, .32}

\usepackage{hyperref}
\usepackage{graphics, appendix, afterpage, makecell} 
\usepackage{bbold}
\usepackage{tikz}
\usepackage{adjustbox}

\usepackage{tcolorbox}

\definecolor{oucrimsonred}{rgb}{0.6, 0.0, 0.0}
\definecolor{persianblue}{rgb}{0.11, 0.22, 0.73}
\definecolor{forestgreen}{rgb}{0.13,0.35,0.13}
\definecolor{lightgray}{rgb}{0.83, 0.83, 0.83}
 \hypersetup{colorlinks, citecolor=oucrimsonred, linkcolor=persianblue, urlcolor=oucrimsonred}
\definecolor{cornellred}{rgb}{0.7, 0.11, 0.11}
\definecolor{navyblue}{rgb}{0.0, 0.0, 0.5}
\definecolor{amethyst}{rgb}{0.6, 0.4, 0.8}
\definecolor{yellow}{rgb}{1.0, 1.0, 0.0}
\definecolor{firebrick}{rgb}{0.7, 0.13, 0.13}
\definecolor{tangerineyellow}{rgb}{1.0, 0.8, 0.0}
\definecolor{deepfuchsia}{rgb}{0.76, 0.33, 0.76}
\definecolor{amber}{rgb}{1.0, 0.75, 0.0}
\definecolor{VioletRed4}{rgb}{0.55, 0.13, .32}
\definecolor{indiagreen}{rgb}{0.07, 0.53, 0.03}
\definecolor{VioletRed4}{rgb}{0.55, 0.13, .32}

\definecolor{oucrimsonred}{rgb}{0.6, 0.0, 0.0}
\newcommand\vertarrowbox[3][6ex]{%
  \begin{array}[t]{@{}c@{}} #2 \\
  \left\uparrow\vcenter{\hrule height #1}\right.\kern-\nulldelimiterspace\\
  \makebox[0pt]{\scriptsize#3}
  \end{array}%
}

\definecolor{mtcolor}{rgb}{.8,.3,.1}

\definecolor{violachiaro}{rgb}{1,0.6,1}

\definecolor{gbcolor}{rgb}{.43,.22,.12}
 
\definecolor{gbcolor2}{rgb}{.9,.2,.6}
\definecolor{gbcolor3}{rgb}{.3,.2,.6}

\definecolor{verdechiaro}{rgb}{0.6,1,0.6}
\definecolor{giallochiaro}{rgb}{1,1,0.6}
\definecolor{bluscuro}{rgb}{0.15, 0.2, 0.9}
\definecolor{verdes}{rgb}{0.1, 0.5, 0.1}%
\definecolor{tangerineyellow}{rgb}{1.0, 0.8, 0.0}
\definecolor{smokyblack}{rgb}{0.06, 0.05, 0.03}

\definecolor{americanrose}{rgb}{1.0, 0.01, 0.24}
\definecolor{cobalt}{rgb}{0.0, 0.28, 0.67}
\definecolor{brandeisblue}{rgb}{0.0, 0.44, 1.0}
\definecolor{mycolor}{rgb}{0.0, 0.0, 0.5}
\definecolor{oxfordblue}{rgb}{0.0, 0.13, 0.28}
\definecolor{azure}{rgb}{0.0, 0.5, 1.0}
\definecolor{turquoiseblue}{rgb}{0.0, 1.0, 0.94}
\newtcolorbox{mynewbox}[1]{colback=white!5!white,colframe=azure!75!black,fonttitle=\bfseries,title=#1}
\newtcolorbox{mybox}{colback=mycolor!5!white,colframe=azure!75!black}
\newtcolorbox{mynamedbox}[1]{colback=mycolor!5!white,colframe=azure!75!black,title=#1}
\definecolor{venetianred}{rgb}{0.78, 0.03, 0.08}
\newtcolorbox{mynamedbox1}[1]{colback=venetianred!5!white,colframe=venetianred!80!black,title=#1}
\newtcolorbox{mynamedbox2}[1]{colback=azure!5!white,colframe=azure!80!black,title=#1}

\definecolor{rossocorsa}{rgb}{0.83, 0.0, 0.0}

\tikzset{->-/.style={decoration={
  markings,
  mark=at position #1 with {\arrow{>}}},postaction={decorate}}}
\tikzset{-<-/.style={decoration={
  markings,
  mark=at position #1 with {\arrow{<}}},postaction={decorate}}} 

\def\be{\begin{equation}}
\def\ee{\end{equation}}
\def\ba{\begin{eqnarray}}
\def\ea{\end{eqnarray}}

\def\L*{{\cal L}_*}
\def\L{\mathcal{L}}
\def\({\left(}
\def\){\right)}

\def\<{\langle}
\def\>{\rangle}

\def\cs2{c_{s}^{2}}

 \def\be   {\begin{equation}}   \def\ee   {\end{equation}}
 \def\ba   {\begin{array}}      \def\ea   {\end{array}}
 \def\bea  {\begin{eqnarray}}   \def\eea  {\end{eqnarray}}
 \def\bean {\begin{eqnarray*}}  \def\eean {\end{eqnarray*}}

\titleclass{\subsubsubsection}{straight}[\subsection]

\newcounter{subsubsubsection}[subsubsection]
\renewcommand\thesubsubsubsection{\thesubsubsection.\arabic{subsubsubsection}}

\titleformat{\subsubsubsection}
  {\normalfont\normalsize\bfseries}{\thesubsubsubsection}{1em}{}
\titlespacing*{\subsubsubsection}
{0pt}{3.25ex plus 1ex minus .2ex}{1.5ex plus .2ex}

\makeatletter
\renewcommand\paragraph{\@startsection{paragraph}{5}{\z@}%
  {3.25ex \@plus1ex \@minus.2ex}%
  {-1em}%
  {\normalfont\normalsize\bfseries}}
\renewcommand\subparagraph{\@startsection{subparagraph}{6}{\parindent}%
  {3.25ex \@plus1ex \@minus .2ex}%
  {-1em}%
  {\normalfont\normalsize\bfseries}}
\def\toclevel@subsubsubsection{4}
\def\toclevel@paragraph{5}
\def\toclevel@paragraph{6}
\def\l@subsubsubsection{\@dottedtocline{4}{7em}{4em}}
\def\l@paragraph{\@dottedtocline{5}{10em}{5em}}
\def\l@subparagraph{\@dottedtocline{6}{14em}{6em}}
\makeatother

\usepackage{titlesec} 
\usepackage[colorlinks,citecolor=blue,urlcolor=blue,hypertexnames=true]{hyperref}
\usepackage{subfigure}

\usepackage[usenames,dvipsnames,svgnames,table]{xcolor}  
\usepackage{tikzsymbols}
\usepackage{natbib}
\usepackage{float}

\usepackage{tikz,xcolor,hyperref}

\usepackage{slashed}

\definecolor{lime}{HTML}{A6CE39}
\DeclareRobustCommand{\orcidicon}{
	\begin{tikzpicture}
	\draw[lime, fill=lime] (0,0) 
	circle [radius=0.2] 
	node[white] {{\fontfamily{qag}\selectfont \tiny ID}};
	\draw[white, fill=white] (-0.0625,0.095) 
	circle [radius=0.007];
	\end{tikzpicture}
	\hspace{-2mm}
}

\foreach \x in {A, ..., Z}{\expandafter\xdef\csname orcid\x\endcsname{\noexpand\href{https://orcid.org/\csname orcidauthor\x\endcsname}
			{\noexpand\orcidicon}}
}

\usepackage{bbold}
\usepackage{tikz}
\usepackage{adjustbox}
\usepackage{tcolorbox}
\usepackage{enumitem}
\usepackage{amsfonts}

\setlist[itemize,1]{label=$\times$}
\setlist[itemize,2]{label=$\checkmark$}
\setlist[itemize,3]{label=$\diamond$}
\setlist[itemize,4]{label=$\bullet$}

\begin{document}
\title{\Large \textcolor{Sepia}{
Primordial non-Gaussianity as a saviour for PBH overproduction in SIGWs  generated by Pulsar Timing Arrays for Galileon inflation
}}
\author{\large Sayantan Choudhury\orcidA{}${}^{1}$}
\email{sayantan\_ccsp@sgtuniversity.org,  \\ sayanphysicsisi@gmail.com (Corresponding author)} 
 \author{\large Kritartha Dey\orcidE{}${}^{1}$}
\email{kritartha09@gmail.com }
\author{\large Ahaskar Karde\orcidB{}${}^{1}$}
\email{kardeahaskar@gmail.com}
\author{\large Sudhakar~Panda\orcidC{}${}^{1,2}$}
\email{panda@niser.ac.in}
\author{\large M.~Sami\orcidD{}${}^{1,3,4}$}
\email{ sami\_ccsp@sgtuniversity.org,  samijamia@gmail.com} 

\affiliation{ ${}^{1}$Centre For Cosmology and Science Popularization (CCSP),\\
        SGT University, Gurugram, Delhi- NCR, Haryana- 122505, India,}
 \affiliation{${}^{2}$School of Physical Sciences,  National Institute of Science Education and Research, Bhubaneswar, Odisha - 752050, India,}
\affiliation{${}^{3}$Center for Theoretical Physics, Eurasian National University, Astana 010008, Kazakhstan.}
	\affiliation{${}^{4}$Chinese Academy of Sciences,52 Sanlihe Rd, Xicheng District, Beijing.}


\begin{abstract}

We investigate the explicit role of negative local non-Gaussianity, $f_{\rm NL}$, in suppressing the abundance of primordial black holes (PBHs) in the single-field model of Galileon inflation. PBH formation requires significantly enhancing the scalar power spectrum, which greatly affects their abundance. The associated frequencies in the nHz regime are also sensitive to the generation of scalar-induced gravitational waves (SIGWs) which may explain the current data from the pulsar timing arrays (PTAs).
Our analysis using the threshold statistics on the compaction function demonstrates that Galileon theory not only avoids PBH overproduction using the curvature perturbation enhancements that give $f_{\rm NL} \sim {\cal O}(-6)$, but also generates SIGWs that conform well with the PTA data.

\end{abstract}

\maketitle
Recent findings from the pulsar timing array (PTA) collaborations, which include the NANOGrav \cite{NANOGrav:2023gor, NANOGrav:2023hde, NANOGrav:2023ctt, NANOGrav:2023hvm, NANOGrav:2023hfp, NANOGrav:2023tcn, NANOGrav:2023pdq, NANOGrav:2023icp}, EPTA \cite{EPTA:2023fyk, EPTA:2023sfo, EPTA:2023akd, EPTA:2023gyr, EPTA:2023xxk, EPTA:2023xiy}, PPTA \cite{Reardon:2023gzh, Reardon:2023zen, Zic:2023gta}, and CPTA \cite{Xu:2023wog}, have confirmed the observed common-spectrum signal for a stochastic gravitational wave background (SGWB). Since then, considerable works have discussed the interpretations and possible cosmological origins of the signal, such as first-order phase transitions, cosmic strings, domain walls, and inflation, particularly scalar-induced gravitational waves (SIGWs). (See refs. \cite{Choudhury:2023hfm,Bhattacharya:2023ysp,Choudhury:2023fjs,Franciolini:2023pbf,Inomata:2023zup,Wang:2023ost,Balaji:2023ehk,HosseiniMansoori:2023mqh,Gorji:2023sil,DeLuca:2023tun,Choudhury:2023kam,Yi:2023mbm,Cai:2023dls,Cai:2023uhc,Huang:2023chx,Vagnozzi:2023lwo,Frosina:2023nxu,Zhu:2023faa,Jiang:2023gfe,Cheung:2023ihl,Oikonomou:2023qfz,Liu:2023pau,Liu:2023ymk,Wang:2023len,Zu:2023olm, Abe:2023yrw, Gouttenoire:2023bqy,Salvio:2023ynn, Xue:2021gyq, Nakai:2020oit, Athron:2023mer,Ben-Dayan:2023lwd, Madge:2023cak,Kitajima:2023cek, Babichev:2023pbf, Zhang:2023nrs, Zeng:2023jut, Ferreira:2022zzo, An:2023idh, Li:2023tdx,Blanco-Pillado:2021ygr,Buchmuller:2021mbb,Ellis:2020ena,Buchmuller:2020lbh,Blasi:2020mfx, Madge:2023cak, Liu:2023pau, Yi:2023npi,Gangopadhyay:2023qjr,Vagnozzi:2020gtf,Benetti:2021uea,Inomata:2023drn,Lozanov:2023rcd,Basilakos:2023jvp,Basilakos:2023xof,Li:2023xtl,Domenech:2021ztg,Yuan:2021qgz,Chen:2019xse,Cang:2023ysz,Cang:2022jyc,Konoplya:2023fmh}). Such cosmological scenarios contain models that better agree with the PTA data. However, this letter focuses on the SIGW interpretation of the PTA signal and concerns related to the primordial black hole (PBH) production associated with the frequency regime of the signal. 

Scalar-induced GWs manifest as the tensor perturbations at second-order induced by curvature perturbations in the spatially flat FLRW background upon horizon re-entry. \cite{Matarrese:1992rp, Matarrese:1993zf, Matarrese:1997ay,Ananda:2006af, Baumann:2007zm}. At the same time, during inflation, small-scale enhancements in the curvature perturbations lead to the generation of overdense and underdense regions in the density contrast field, which, on exceeding a threshold overdensity, gravitationally collapses to form PBHs \cite{Zeldovich:1967lct,Hawking:1974rv,Carr:1974nx,Carr:1975qj,Chapline:1975ojl,Carr:1993aq,Choudhury:2012whm,Choudhury:2012yh,Choudhury:2013woa,Yokoyama:1998pt,Kawasaki:1998vx,Rubin:2001yw,Khlopov:2002yi,Khlopov:2004sc,Saito:2008em,Khlopov:2008qy,Carr:2009jm,Choudhury:2011jt,Lyth:2011kj,Drees:2011yz,Drees:2011hb,Ezquiaga:2017fvi,Kannike:2017bxn,Hertzberg:2017dkh,Pi:2017gih,Gao:2018pvq,Dalianis:2018frf,Cicoli:2018asa,Ozsoy:2018flq,Byrnes:2018txb,Ballesteros:2018wlw,Belotsky:2018wph,Martin:2019nuw,Ezquiaga:2019ftu,Motohashi:2019rhu,Fu:2019ttf,Ashoorioon:2019xqc,Auclair:2020csm,Vennin:2020kng,Nanopoulos:2020nnh,Inomata:2021uqj,Stamou:2021qdk,Ng:2021hll,Wang:2021kbh,Kawai:2021edk,Solbi:2021rse,Ballesteros:2021fsp,Rigopoulos:2021nhv,Animali:2022otk,Correa:2022ngq,Frolovsky:2022ewg,Escriva:2022duf,Ozsoy:2023ryl,Ivanov:1994pa,Afshordi:2003zb,Frampton:2010sw,Carr:2016drx,Kawasaki:2016pql,Inomata:2017okj,Espinosa:2017sgp,Ballesteros:2017fsr,Sasaki:2018dmp,Ballesteros:2019hus,Dalianis:2019asr,Cheong:2019vzl,Green:2020jor,Carr:2020xqk,Ballesteros:2020qam,Carr:2020gox,Ozsoy:2020kat,Baumann:2007zm,Saito:2008jc,Saito:2009jt,Choudhury:2013woa,Sasaki:2016jop,Raidal:2017mfl,Papanikolaou:2020qtd,Ali-Haimoud:2017rtz,Di:2017ndc,Raidal:2018bbj,Cheng:2018yyr,Vaskonen:2019jpv,Drees:2019xpp,Hall:2020daa,Ballesteros:2020qam,Carr:2020gox,Ozsoy:2020kat,Ashoorioon:2020hln,Papanikolaou:2020qtd,Wu:2021zta,Kimura:2021sqz,Solbi:2021wbo,Teimoori:2021pte,Cicoli:2022sih,Ashoorioon:2022raz,Papanikolaou:2022chm,Papanikolaou:2023crz,Wang:2022nml,ZhengRuiFeng:2021zoz,Cohen:2022clv,Arya:2019wck,Correa:2022ngq,Cicoli:2022sih,Brown:2017osf,Palma:2020ejf,Geller:2022nkr,Braglia:2022phb,Frolovsky:2023xid,Aldabergenov:2023yrk,Aoki:2022bvj,Frolovsky:2022qpg,Aldabergenov:2022rfc,Ishikawa:2021xya,Gundhi:2020kzm,Aldabergenov:2020bpt,Cai:2018dig,Cheng:2021lif,Balaji:2022rsy,Qin:2023lgo,Riotto:2023hoz,Ragavendra:2020vud,Ragavendra:2021qdu,Ragavendra:2020sop,Gangopadhyay:2021kmf,Papanikolaou:2022did, Choudhury:2024jlz, Choudhury:2024dei, Choudhury:2024ybk, Choudhury:2024aji}. However, in some cases, PBHs may also get overproduced \cite{Franciolini:2023pbf,Ferrante:2022mui,DeLuca:2023tun,Gorji:2023sil,Gorji:2023ziy}. 
The general mechanism for PBH production in single-field inflationary models necessitates an ultra-slow roll (USR) phase, which sufficiently amplifies the scalar power spectrum amplitude \cite{Kristiano:2022maq,Riotto:2023hoz,Banerjee:2021lqu,Choudhury:2023vuj,Choudhury:2023jlt,Choudhury:2023rks,Choudhury:2023hvf,Choudhury:2023kdb,Choudhury:2023hfm,Kristiano:2023scm,Riotto:2023gpm,Firouzjahi:2023ahg,Firouzjahi:2023aum,Franciolini:2023lgy,Cheng:2023ikq,Tasinato:2023ukp,Motohashi:2023syh,Choudhury:2024aji}. This USR phase raises concerns about generating large non-Gaussianities (NGs) \cite{Choudhury:2023kdb}, challenging the  Gaussian statistics assumption for the comoving curvature perturbations. Accurately assessing the PBH abundance, as the fraction constituting all of present-day dark matter, requires a careful account of the primordial NGs.  
Recent attempts to tackle PBH overproduction involve strategies like incorporating an 
extra spectator field along with the metric perturbations, as in the curvaton model \cite{Ferrante:2023bgz,Franciolini:2023pbf,Gow:2023zzp} or an additional spectator tensor field \cite{Gorji:2023sil}, while positive NG is declared harmful to the PBH abundance, especially in non-attractor single-field models \cite{Firouzjahi:2023xke}. 
The total PBH abundance heavily depends on the NG magnitude and its signature. Requiring a sizeable abundance largely impacts the peak of the scalar power spectrum, thereby putting constraints on the maximum of the observed SIGW spectrum.

In this letter, we aim to demonstrate that PBH overproduction does not occur in the non-attractor single-field inflation model described by covariantized Galileon theory, which includes a USR phase while accounting for the primordial NGs and intrinsic non-linearities between the density fluctuations and the comoving curvature perturbations.

The study by authors in ref.\cite{Choudhury:2023kdb} demonstrates significant negative NG, $f_{\rm NL} \sim {\cal O}(-6)$, generated in the USR phase of Galileon theory. Its unique non-renormalization theorem strongly suggests suppression of the quantum loop corrections within the theory, allowing no constraints on the allowed PBH masses \cite{Choudhury:2023kdb,Choudhury:2023hfm,Choudhury:2023hvf}. This property also provides us with the advantage of allowing for successful inflation.
Our findings from Galileon reveal that large negative NGs allow for a sizeable abundance of PBH and a substantial scalar power spectrum amplitude, consistent with the recent NANOGrav 15 data. Another reason, observationally speaking, for working with Galileon comes from the results in \cite{Choudhury:2023hfm}, where the SIGW spectrum generated shows consistency with the NANOGrav 15 signal and, thanks to the intrinsic features of this theory, it also shows signatures in the parameter space of the existing and future GW experiments, including LISA \cite{LISA:2017pwj}, DECIGO \cite{Kawamura:2011zz}, ET \cite{Punturo:2010zz}, CE \cite{Reitze:2019iox}, BBO \cite{Crowder:2005nr}, HLVK \cite{LIGOScientific:2014pky, VIRGO:2014yos, KAGRA:2018plz}, and HLV(O3) \cite{LIGOScientific:2014pky, VIRGO:2014yos, KAGRA:2018plz}.

We employ the approach of threshold statistics on the compaction function parameter to accurately investigate the PBH abundance \cite{Ferrante:2022mui,Franciolini:2023pbf,Musco:2020jjb,Kehagias:2019eil,Young:2019yug,Musco:2018rwt,Gow:2022jfb,Biagetti:2021eep}. This procedure helps us to picture the non-linear corrections to the density contrast and include the primordial NGs in the curvature perturbation within the framework of covariantized Galileon theory. 

\textcolor{black}{We start by briefly laying out the underlying features of the Galileon inflation theory \cite{Burrage:2010cu} used in this paper. A remarkable quality of this theory is that despite the action containing a higher-derivative structure one obtains equations of motion quadratic in the scalar field $\phi$ which is known here as the Galileon field. The theory comes equipped with a unique symmetry known as the Galilean shift symmetry, which in a $3+1$ spacetime can be written as follows:}
\bea \label{galsym}
\phi \rightarrow \phi + v_{\mu}x^{\mu} + b,
\eea
\textcolor{black}{where $b$ is a constant scalar, $v^{\mu}$ is a constant vector, $x^{\mu}$ represent the spacetime variable. We will return to the use of this symmetry when we talk about the importance of the non-renormalization theorem and quantum loop effects in Galileon theory.
We mostly focus on the dynamics of this Galileon field in a inflationary scenario where the background spacetime has a quasi de-Sitter geometry, such that the effective potential during inflation satisfies the condition $|\Delta V/V|\ll 1$ and the expansion is characterized by the necessary slow-roll conditions. The action for the Galileon field
in the background, $\bar{\phi}(t)$ reads:}
\bea \label{action}  S^{(0)}&=&\int d^4x\,a^3 \,\Bigg[\dot{\bar{\phi}}^2\Bigg\{\frac{c_2}{2}+2c_3Z+\frac{9c_4}{2}Z^2+6c_5Z^3\Bigg\}\nonumber\\ &+& c_1\bar{\phi}\Bigg],
\quad \text{where}\quad Z \equiv H\dot{\bar{\phi}}/\Lambda^3, \quad c_{1}=\lambda^{3},
\eea
\textcolor{black}{with $c_{i}\forall\;i=1,\cdots,5$ representing the coefficients that parameterize the Galileon theory, $\lambda$ is a mass dimension $1$ parameter that also helps to characterize the exact shift symmetry breaking in the potential, and the action involves the coupling parameter, $Z$, in terms of the physical cut-off scale, $\Lambda$, of the theory. From here the task of computing the scalar power spectrum involves expanding this action to second order in the perturbations around the background, more details on this construction and the comoving curvature perturbation can be found in ref.\cite{Choudhury:2023hvf}. This perturbed second-order action in the Fourier space reads as follows:}
\bea \label{quadaction}
S^{(2)}_{\zeta} &=& \int d\tau\;\frac{d^{3}{\bf k}}{(2\pi)^{3}}a^{2}(\tau)\frac{\cal A}{H^{2}}\left(|\zeta_{{\bf k}}^{'}(\tau)|^{2} - c_{s}^{2}k^{2}|\zeta_{{\bf k}}(\tau)|^{2}\right), \nonumber\\
&&\quad\quad\quad\quad \text{where}\quad c_{s}^{2} = \frac{{\cal B}}{{\cal A}},
\eea
\textcolor{black}{here $\zeta_{\bf k}(\tau)$ is the Fourier space comoving curvature perturbation mode function, $a(\tau)=-1/H\tau$ is the scale factor in conformal time coordinates and, ${\cal A},{\cal B}$, are the time-dependent coefficients having relation with $c_{s}$ as the effective sound speed parameter of our theory. These coefficients have the following specific expressions:}
\bea \label{coeffA}
    {\cal A}&\equiv& \frac{\dot{\bar{\phi}}^2}{2}\Bigg(c_2+12c_3Z+54c_4Z^2+120c_5Z^3\Bigg),\\
    \label{coeffB} {\cal B}&\equiv& 
   \frac{\dot{\bar{\phi}}^2}{2}\Bigg\{c_2+4c_3\Bigg(2Z-\frac{H\dot{\bar{\phi}}}{\Lambda^3}\eta\Bigg)+2c_4\Bigg[13Z^2-\frac{6}{\Lambda^6}\nonumber\\
   &&\quad\times \dot{\bar{\phi}}^2H^2\big(\epsilon+2\eta\big)\Bigg]-\frac{24c_5}{\Lambda^9}H^3\dot{\bar{\phi}}^3\big(2\epsilon+1\big)\Bigg\},
\eea
\textcolor{black}{and they also include the first and second slow-roll parameters, $\epsilon=-\dot{H}/H^{2}$ and $\eta=\dot{\epsilon}/\epsilon H$, respectively.} Important to note is the fact that the values of the dimensionless coefficients, $\{c_{i}\;\forall i=1\; \text{to}\;5 \}$, and the coupling $Z$ and field $\bar{\phi}(t)$, are chosen in such a manner to allow for the parameter $c_{s}$ taking proper values that satisfy causality and unitarity constraints during inflation. Also, the chosen values help to set up an additional ultra-slow roll phase for the brief duration of $\Delta{\cal N}_{\rm USR} \sim {\cal O}(2)$ in the present theory \cite{Choudhury:2024one}. 
\textcolor{black}{We hereafter aim to consider the scalar power spectrum from Galileon theory, accounting for the quantum one-loop corrections originating due to the perturbed action at cubic order. The complete construction of such an action and dealing with the quantum loop effects requires understanding the significance of the non-renormalization theorem to which we now allude. }

\textcolor{black}{The non-renormalization theorem provides a great deal of simplicity in the computation of one-loop effects in Galileon theory as it allows us to completely do away with operators that bring significant radiative corrections to its Lagrangian. Since we focus on an inflationary scenario from the start, a mild shift symmetry breaking is needed to realise the same physically. However, under such a soft symmetry-breaking condition, the non-renormalization theorem remains valid since the Galileon theory operators, along with any couplings of the heavy fields to the Galileon, respect the Galilean shift symmetry. This feature protects the Lagrangian against any harmful renormalizations that destroy its stability \cite{Burrage:2010cu,Goon:2016ihr}. Consider the Galilean symmetry in eqn.(\ref{galsym}). Under this symmetry the comoving curvature perturbation transform in the following manner:}
\bea
\zeta \rightarrow \zeta - \frac{H}{\dot{\bar{\phi}}}v.\delta x
\eea
\textcolor{black}{from which we can deduce that the terms $\zeta$, $\zeta'$, and $\partial_{i}\zeta$, all break Galilean shift symmetry mildly, and only $\partial^{2}\zeta$ remains invariant due to its double derivative structure. Using the above transformation properties one can construct operators as a combination of these terms that exhibit the mild symmetry breaking needed for inflation. After going through the algebra of removing any redundant contributions via field re-definitions and other boundary terms, we remain with only a few acceptable terms such as, $\zeta'^{3}$, $\zeta'^{2}\partial^{2}\zeta$, $\zeta'(\partial_{i}\zeta)^{2}$ and $(\partial_{i}\zeta)^{2}\partial^{2}\zeta$, which later form the cubic order action.  We now provide a glimpse of this said cubic action that is responsible for the quantum loop effects calculations: }

\begin{widetext}
\begin{eqnarray}
\label{cubicaction}
S^{3}_{\zeta} = \int d\tau\; d^{3}x\frac{a(\tau)^{2}}{H^{3}}\bigg[\frac{{\cal G}_1}{a}\zeta^{'3}+\frac{{\cal G}_2}{a^2}\zeta^{'2}\left(\partial^2\zeta\right)+\frac{{\cal G}_3}{a}\zeta^{'}\left(\partial_i\zeta\right)^2+\frac{{\cal G}_4}{a^2}\left(\partial_i\zeta\right)^2\left(\partial^2\zeta\right)\bigg].
\end{eqnarray}
\textcolor{black}{We do not list here explicitly the Galileon operators coefficients, ${\cal G}_{i}\;\forall i=1,\cdots,4$, due to their complex structure and would be irrelevant, but their exact expressions can be found in \cite{Choudhury:2023hvf}. The actual one-loop computation requires following the Schwinger-Keldysh (\textit{in-in}) formalism, and its implementation results in the two-point correlation function read as follows:}
\begin{eqnarray}
\langle\hat{\zeta}_{\bf k}\hat{\zeta}_{-{\bf k}}\rangle&=&\lim_{\tau\rightarrow 0}\left\langle\bigg[\overline{T}\exp\bigg(i\int^{\tau}_{-\infty(1-i\epsilon)}d\tau^{'}\;H_{\rm int}(\tau^{'})\bigg)\bigg]\;\;\hat{\zeta}_{\bf k}(\tau)\hat{\zeta}_{-{\bf k}}(\tau)\;\;\bigg[{T}\exp\bigg(-i\int^{\tau}_{-\infty(1+i\epsilon)}d\tau^{''}\;H_{\rm int}(\tau^{''})\bigg)\bigg]\right\rangle,
\end{eqnarray}
\textcolor{black}{where inside the above expression, we expand the interaction Hamiltonian and upto leading order obtain the relevant one-loop contributions after plugging inside, $H_{\rm int}=-{\cal L}^{3}$, using the action in eqn. (\ref{cubicaction}). This identification of $H_{\rm int}$ works specifically at the third order. All that now remains is to calculate the following in-in correlations:}
\begin{eqnarray}
    &&\label{tree}\langle\hat{\zeta}_{\bf k}\hat{\zeta}_{-{\bf k}}\rangle_{(0,0)}=\langle\hat{\zeta}_{\bf k}\hat{\zeta}_{-{\bf k}}\rangle_{\bf Tree}=\lim_{\tau\rightarrow 0}\langle \hat{\zeta}_{\bf k}(\tau)\hat{\zeta}_{-{\bf k}}(\tau)\rangle,\\
    &&\label{loop1}\langle\hat{\zeta}_{\bf k}\hat{\zeta}_{-{\bf k}}\rangle_{(0,2)}=\lim_{\tau\rightarrow 0}\int^{\tau}_{-\infty}d\tau_1\;\int^{\tau}_{-\infty}d\tau_2\;\langle \hat{\zeta}_{\bf k}(\tau)\hat{\zeta}_{-{\bf k}}(\tau)H_{\rm int}(\tau_1)H_{\rm int}(\tau_2)\rangle,\\
  &&\label{loop2}\langle\hat{\zeta}_{\bf k}\hat{\zeta}_{-{\bf k}}\rangle^{\dagger}_{(0,2)}=\lim_{\tau\rightarrow 0}\int^{\tau}_{-\infty}d\tau_1\;\int^{\tau}_{-\infty}d\tau_2\;\langle \hat{\zeta}_{\bf k}(\tau)\hat{\zeta}_{-{\bf k}}(\tau)H_{\rm int}(\tau_1)H_{\rm int}(\tau_2)\rangle^{\dagger},\\
  &&\label{loop3}\langle\hat{\zeta}_{\bf k}\hat{\zeta}_{-{\bf k}}\rangle_{(1,1)}=\lim_{\tau\rightarrow 0}\int^{\tau}_{-\infty}d\tau_1\;\int^{\tau}_{-\infty}d\tau_2\;\langle H_{\rm int}(\tau_1)\hat{\zeta}_{\bf k}(\tau)\hat{\zeta}_{-{\bf k}}(\tau)H_{\rm int}(\tau_2)\rangle,\\
  &&\label{totalloop}\langle\hat{\zeta}_{\bf k}\hat{\zeta}_{-{\bf k}}\rangle_{\bf One-loop}= \langle\hat{\zeta}_{\bf k}\hat{\zeta}_{-{\bf k}}\rangle_{(0,2)} + \langle\hat{\zeta}_{\bf k}\hat{\zeta}_{-{\bf k}}\rangle^{\dagger}_{(0,2)} + \langle\hat{\zeta}_{\bf k}\hat{\zeta}_{-{\bf k}}\rangle_{(1,1)}.
\end{eqnarray}
\end{widetext}
\textcolor{black}{The correlation in eqn. (\ref{tree}) provides the tree-level contribution to the primordial scalar power spectrum while the sum of the correlations in eqns. (\ref{loop1}-\ref{loop3}) provide the necessary one-loop corrections written in eqn. (\ref{totalloop}). The final power spectrum expression combines the contributions coming from each of the four operators inside eqn. (\ref{cubicaction}) and thus leads to the result as: }
\bea\label{finalcorrl}\langle\hat{\zeta}_{\bf k}\hat{\zeta}_{-{\bf k}}\rangle &=&\langle\hat{\zeta}_{\bf k}\hat{\zeta}_{-{\bf k}}\rangle_{\bf Tree}+\langle\hat{\zeta}_{\bf k}\hat{\zeta}_{-{\bf k}}\rangle_{\bf One-loop},\eea
\textcolor{black}{where the one-loop effects remain highly suppressed when calculated as described above due to the non-renormalization theorem. We however provide the necessary details as it will be helpful to visualize the total power spectrum structure in the upcoming discussions. }

The power spectrum associated with the scalar modes is crucial to analyze the SIGW spectrum. Also, the enhancements in the scalar power spectrum of our theory due to the presence of the USR phase will give rise to a peak amplitude corresponding to the maximum abundance of PBHs. We begin by presenting the final cut-off regularized one-loop corrected power spectrum for the scalar modes as follows:
     \bea
     \label{powerspec}
       \Delta^{2}_{\zeta}(k) &=& A \bigg[1+\left(\frac{k}{k_s}\right)^{2}\bigg]\nonumber\\
       &&\times \bigg\{\left(\frac{k_s}{k_e}\right)^{6}\left(1+{\cal Q}_c\right)+\bigg(\big|\alpha_{\bf k}^{(2)}-\beta_{\bf k}^{(2)}\big|^2 \;\Theta(k-k_s)\nonumber\\
       &&+\big|\alpha_{\bf k}^{(3)}-\beta_{\bf k}^{(3)}\big|^2 \;\Theta(k-k_e)\bigg)\bigg\},
    \eea

    \begin{figure*}[htb!]
    	\centering
   {
   \includegraphics[width=18.5cm,height=7cm] {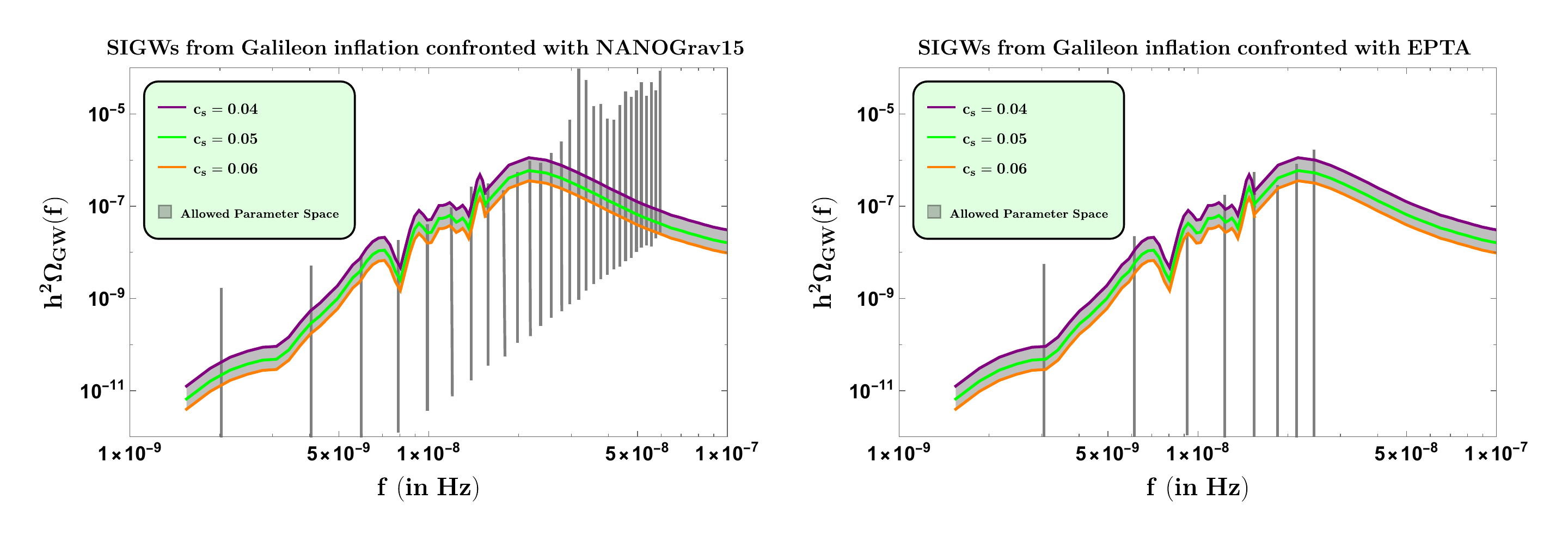}
    } 
    	\caption[Optional caption for list of figures]{The SIGW spectra from Galileon theory confronted with the NANOGrav 15 signal (\textit{left-panel}), and with the EPTA signal (\textit{right-panel}). The gray-colored band illustrates the permissible parameter space for various values of the effective sound speed at the pivot scale, $c_{s} \in \{0.04,0.05,0.06\}$, ensuring $f_{\rm NL} \sim {\cal O}(-6)$. The gray lines in the background show the NANOGrav 15 and EPTA signal.} 
    	\label{nanoeptafig}
    \end{figure*}
where ${\cal Q}_c$ denotes the one-loop corrections present in the final scalar power spectrum, which becomes suppressed on account of the non-renormalization theorem in the Galileon theory. \textcolor{black}{We refer the readers to \cite{Choudhury:2023hvf} for the explicit computations of the loop effects contributions ${\cal Q}_c$ resulting from the use of eqns. (\ref{loop1}-\ref{loop3}).} Also $k_s/k_e\sim {\cal O}(0.1)<1$.
The quantity $A$ represents the amplitude of the scalar power spectrum at the scale $k_s$, and  $\{k_{s},\;k_{e}\}$ denote the transition scales from the first slow-roll (SRI) to the USR and from the USR to the second slow-roll (SRII) phases, respectively. \textcolor{black}{The amplitude $A$ involves the coefficients defined in eqns.(\ref{coeffA},\ref{coeffB}) and the effective sound speed $c_{s}$ from eqn.(\ref{action}). For details regarding the comoving curvature perturbation mode solutions and the scalar power spectrum with Bogoliubov coefficients $\alpha^{(2)}_{\bf k},\;\beta^{(2)}_{\bf k},\;\alpha^{(3)}_{\bf k},\;\beta^{(3)}_{\bf k}$, see the appendix \ref{App:A}.}
The transition scale, $k_{s}$, governs the type of PBH produced within the range of sub-solar to massive solar mass.

The fractional energy density in the currently observed GW spectrum is measured using the total scalar power spectrum from the relation \cite{Kohri:2018awv}:
\bea \label{GWspec}
\Omega_{\rm GW}(f) &=& 0.39\times\left[\frac{g_{*}(T_{\rm rad})}{106.75}\right]^{-1/3}\Omega_{r,0}\times \int^{\infty}_{0}dy \nonumber\\
&& \int^{1+y}_{|1-y|}dx\;{\cal G}(x,y)\Delta^{2}_{\zeta}(kx)\Delta^{2}_{\zeta}(ky),
\eea
where the present radiation energy density fraction is denoted by $\Omega_{r,0}$, and $g_{*}(T_{\rm rad})$ is the relativistic d.o.f during radiation-dominated (RD) era at the specific temperature $T_{\rm rad}$. The integration kernel is used as: 
\bea
{\cal G}(x,y) &=& \frac{3\;\big(4x^2 -(1+y^2 -x^2)^2\big)^2\;\big(x^2+y^2-3\big)^4}{1024\;x^8y^8}\nonumber\\
&& \times \bigg[\bigg(\ln \frac{|3-(x+y)^2|}{|3-(x-y)^2|} -\frac{4xy}{x^2 +y^2 -3}\bigg)^2\nonumber\\
&&+ \;\pi^2 \Theta(x+y-\sqrt{3})\bigg].
\eea 
The relation to convert between the frequency and the wavenumber is used here as:
\bea
f = 1.6 \times 10^{-9}\;{\rm Hz}\left(\frac{k}{10^{6}{\rm Mpc}^{-1}}\right).
\eea

The eqn.(\ref{GWspec}) contains the predominant contribution to the GW spectrum coming from the RD era and behaviour of the peak amplitude of the scalar power spectrum, ${\cal O}(10^{-2})$, 
where the impact of the corrections due to NGs to the curvature perturbation is subdominant. Our task would be to use the total scalar power spectrum as in eqn.(\ref{powerspec}) to determine the SIGW spectrum within the frequency range sensitive to the NANOGrav 15, as well as the EPTA signal, where we would also examine the production of the PBHs. 

Fig.(\ref{nanoeptafig}) presents our results for the SIGW spectra from Galileon theory superimposed over the NANOGrav 15 and EPTA signal. We plot the spectra for a set of effective sound speed values, $c_{s} \in \{0.04,0.05,0.06\}$, where the condition $f_{\rm NL} \sim {\cal O}(-6)$ is satisfied, as shown in \cite{Choudhury:2023kdb}. The generated spectra from Galileon theory lie within the observed peak and low-frequency regime of the signals. For the full range of both the signals, the actual trend is obeyed largely by the resulting spectra, which achieves a peak value during the interval of the USR phase where $f \sim {\cal O}(10^{-8}-10^{-7}){\rm Hz}$.

\begin{figure}[htb!]
    	\centering
   {
   \includegraphics[width=9cm,height=7cm] {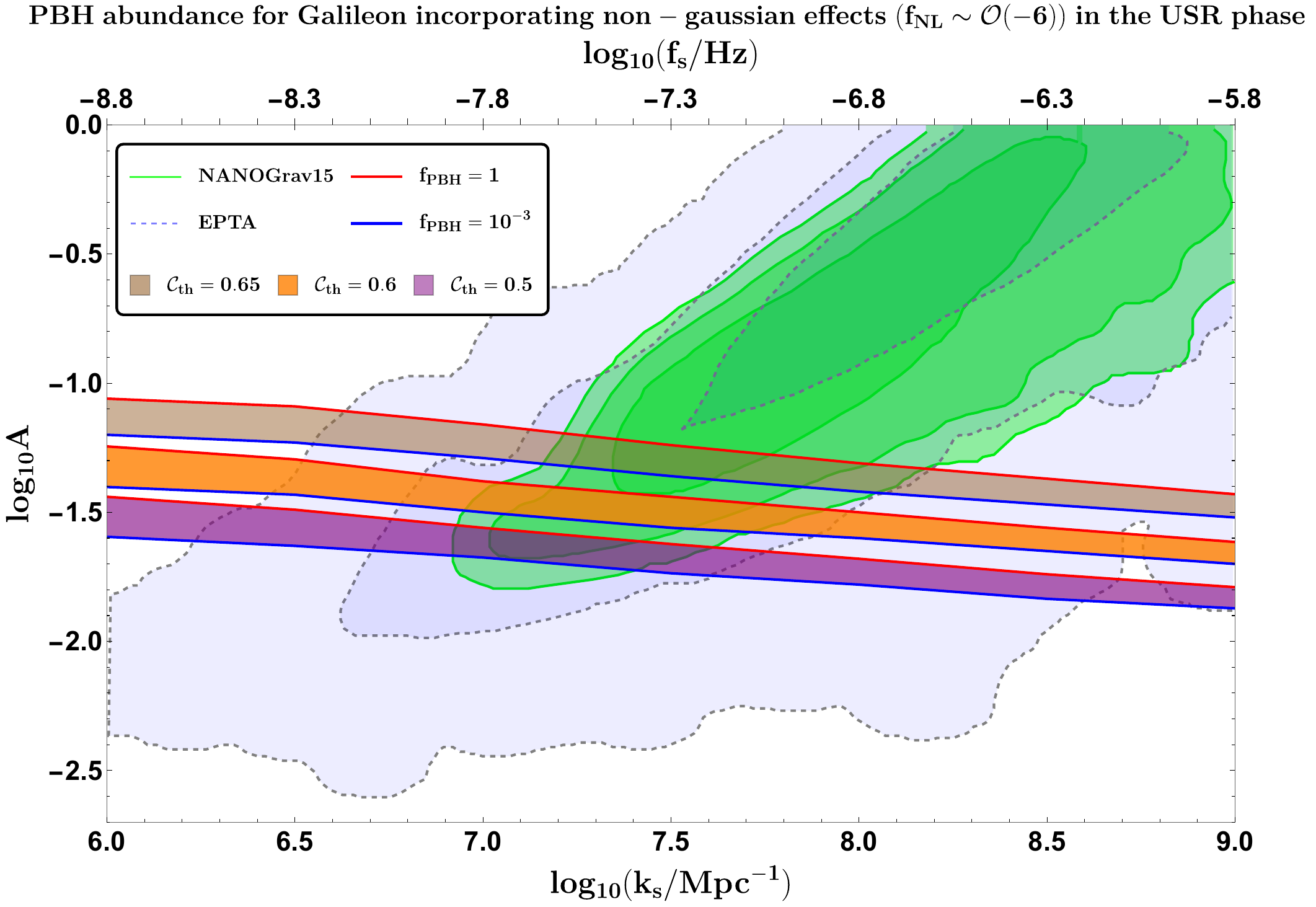}
    }
    	\caption[Optional caption for list of figures]{PBH abundance from Galileon theory for different values of the volume-averaged density contrast threshold. The brown, orange, and purple colored bands correspond to the threshold values in $\mathcal{C}_{\rm th}=\{0.65,0.6,0.5\}$, respectively. The red and blue borders of both bands bound the region of abundance $f_{\rm PBH} \in (1, 10^{-3})$. The green and light-blue posteriors representing, respectively, the NANOGrav 15 and EPTA are taken from \cite{Franciolini:2023pbf}.} 
    	\label{pbhfig}
    \end{figure}

We now address the central issue concerning the PBH abundance and related significant overproduction issues. An accurate estimation of the abundance requires properly considering the non-linearities between the perturbations in the density contrast field and the comoving curvature given by \cite{Musco:2018rwt}:
\bea \label{NLrelation}  \delta(r, t) &=& -\frac{2}{3}\frac{u}{(aH)^2}e^{-2\zeta(r)} \bigg[\zeta^{''}(r) + \frac{2}{r}\zeta^{'}(r) + \frac{1}{2}\zeta^{'}(r)^{2}\bigg],\quad\quad \eea
where $u = 3(1+w)/(5+3w)$ is related to the equation of state parameter $w$ \cite{Polnarev:2006aa}. We will further use $w=1/3$ for the RD era in the rest of computation. 
The above equation assumes spherical symmetry for the locally perturbed region and the long wavelength approximation on super-horizon scales. The further inclusion of primordial NGs in the conserved curvature perturbation random field $\zeta(r)$ will then translate into the density contrast field through the above-mentioned non-linear relation. We focus on the quadratic NG model in this work, given by the well-known ansatz:
\bea \zeta = \zeta_{G} + \frac{3}{5}f_{\rm NL}\zeta_{G}^{2}, \eea
where $\zeta_{G} \equiv \zeta_{G}(r)$ follows Gaussian statistics and $f_{\rm NL}$ measures the amount of NG in the theory. We now invoke the threshold statistics approach on the compaction function to accurately estimate PBH abundance in the presence of the above-mentioned features. Details concerning the compaction function can be found in the SM.  From the use of eqn.(\ref{NLrelation}), the definition of the compaction function ${\cal C}(r,t)$ then acquires a time-independent behavior valid in the super-horizon scales, resulting in the following expression:
\bea
\label{Cg}
{\cal C}(r) = {\cal C}_{G}(r)\frac{d\zeta}{d\zeta_{G}} - \frac{1}{4 u}\left({\cal C}_{G}(r)\frac{d\zeta}{d\zeta_{G}}\right)^{2},
\eea 
where ${\cal C}_{G}(r) = -2ur\zeta^{'}_{G}(r)$. The above relation contains two Gaussian variables: $\zeta_{G}(r)$ and ${\cal C}_{G}(r)$, defined as the derivative of a Gaussian variable. Now, the total fraction of the dark matter present in PBHs comes from after integrating over a range of horizon masses ($M_{H}$) in the relation \cite{Ferrante:2022mui, Franciolini:2023pbf}: 
\bea \label{fpbh} f_{\rm PBH} &=& \frac{1}{\Omega_{\rm DM}}\int d\;\ln{M_{H}}\left(\frac{M_{H}}{M_{\odot}}\right)^{-\frac{1}{2}}\nonumber\\
&& \times\left(\frac{g_{*}}{106.75}\right)^{\frac{3}{4}}\left(\frac{g_{*s}}{106.75}\right)^{-1}\left(\frac{\beta_{\rm NG}(M_{H})}{7.9 \times 10^{-10}}\right),
\eea
    \begin{figure*}[htb!]
    	\centering
   { 
   \includegraphics[width=18cm,height=6cm] {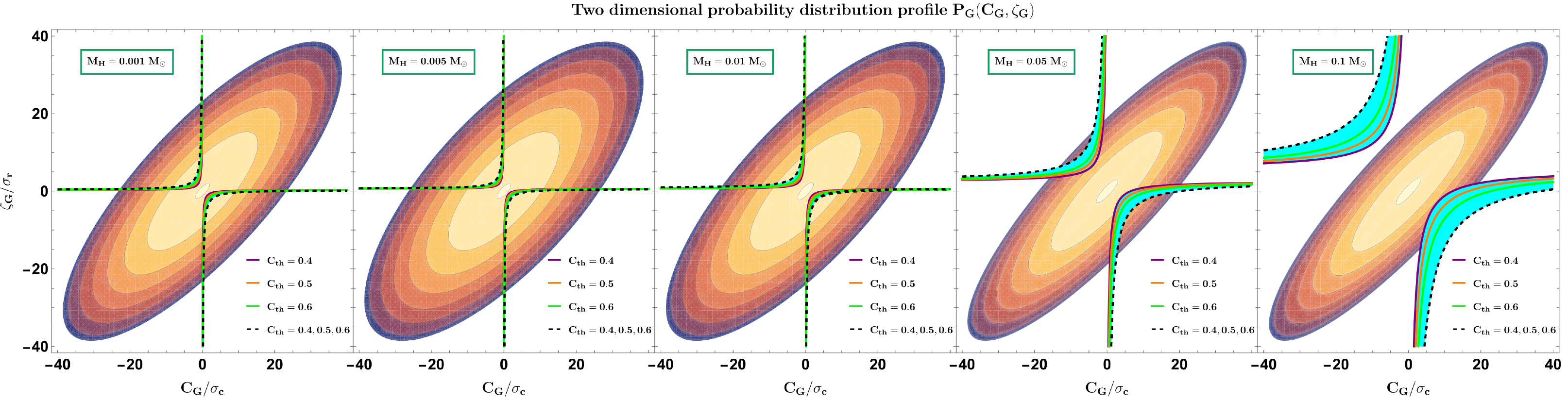}
    }
 	\caption[Optional caption for list of figures]{ Logarithmic plot of 2D PDF for various horizon mass $M_H$, including multiple values of the threshold density contrast, namely ${\cal C}_{\rm th}={0.4,0.5,0.6}$. $\mathds{P}_G ({\cal C}_G,{\cal \zeta}_G)$ is indicated by the contour lines plotted against the Gaussian variables: ${\cal C}_G,{\cal \zeta}_G$. The shaded region between the two dotted lines indicates the parameter space, representing the integration domain to obtain a sizeable PBH mass fraction. All the plots are obtained for the scalar power spectrum having an amplitude $A=10^{-2}$.} 
    	\label{2dpdffig}
    \end{figure*}
where $\Omega_{\rm DM} \simeq 0.264$ represent the dark matter density of the universe and $g_{*},\;g_{*s}$ represent the effective energy and entropy degrees of freedom. 
It is well-known that the formation of PBH shows exponential sensitivity to the tail of the Probability Distribution Function (PDF) of the density fluctuations where the non-Gaussian effects are prominent \cite{DeLuca:2022rfz,Taoso:2021uvl,Atal:2018neu,Young:2013oia,Byrnes:2012yx,Bullock:1996at}. In the present scenario, there exist non-zero auto and cross-correlations between the two Gaussian random variables, $\cal{C}_G$ and $\zeta_G$, which lead to the two-dimensional joint PDF:
 \vspace{-0.2cm}
   \bea
   \label{PDF}
 \mathds{P}_G({\cal C}_G,\zeta_G) &=& \frac{1}{2\pi\sigma_c \sigma_{r}\sqrt{1-\gamma_{\rm cr}^2}}\exp{\left(-\frac{\zeta_G ^2}{2\sigma_r ^2}\right)}\nonumber\\
  && \times \exp{\bigg[\frac{-1}{2(1-\gamma_{\rm cr}^2)}\bigg(\frac{{\cal C}_G}{\sigma_c}-\frac{\gamma_{\rm cr}\zeta_G}{\sigma_r}\bigg)^2\bigg]},
   \eea
with $\gamma_{cr}=\sigma^{2}_{cr}/(\sigma_{c}\sigma_{r})$ representing the correlation coefficient. This PDF helps to calculate the required mass fraction of PBHs when considering the domain set by the threshold statistics on the compaction function (for details, see appendix \ref{App:B}) \cite{Ferrante:2022mui}:
\bea   
\label{Beta}
\beta_{\rm NG}(M_{H}) = \int_{\cal D}{\cal K}({\cal C}-{\cal C}_{\rm th})^{\gamma}\mathds{P}_{G}({\cal C}_G, \zeta_G) d{\cal C}_G\;d\zeta_{G}, \eea
where the \textit{critical scaling relation} is incorporated through ${\cal K}({\cal C}-{\cal C}_{\rm th})^{\gamma}$ for the PBH mass formed during horizon re-entry \cite{Choptuik:1992jv, Evans:1994pj}. The values for constant ${\cal K}\sim {\cal O}(1-10)$ and the threshold ${\cal C}_{\rm th}$, 
based on \cite{Musco:2020jjb}, arises from the simulations, and $\gamma \sim 0.36$ for the RD era. The integration domain is given by ${\cal D} = \{{\cal C}(r) \geq {\cal C}_{\rm th} \wedge {\cal C}_{G}(d\zeta/d\zeta_{G}) \leq 2u\}$, which involves maximizing the compaction function, the details for which we will provide in the SM.
The 2D joint PDF contains various correlations for which we now provide the definitions \cite{Franciolini:2023pbf,Young:2022phe}:
\bea \label{sigcr}
\sigma_{cr}^{2} &=& \frac{2u}{3} \int_{0}^{\infty}\frac{dk}{k}(k r_{m})^{2}W_{g}(k,r_{m})W_{s}(k,r_{m})\tilde{\Delta}^{2}_{\zeta}(k),\quad
\\
\label{sigc} \sigma_{c}^{2} &=& \left(\frac{2u}{3}\right)^{2}\int_{0}^{\infty}\frac{dk}{k}(k r_{m})^{4}W_{g}^{2}(k,r_{m})\tilde{\Delta}^{2}_{\zeta}(k),
\\
\label{sigr} \sigma_{r}^{2} &=& \int_{0}^{\infty}\frac{dk}{k}W^{2}_{s}(k,r_{m})\tilde{\Delta}^{2}_{\zeta}(k),
\eea
\textcolor{black}{where we choose the smoothing functions, $W_{g}(k,r)$ and $W_{s}(k,r)$, to be Gaussian in nature, i.e., $\exp{(-k^{2}r^{2}/2)}$. The function $W_{s}(k,r)$ is referred to in the literature as the spherical-shell window function. For our purposes, we found that the choice of both such window functions to be Gaussian works well, rather than a top-hat and $W_{s}(k,r)=\sin(kr)/kr$ structure. This approach effectively smoothens small-scale fluctuations, and we will later observe its impact when evaluating the abundance of PBH in fig.(\ref{pbhfig}).}
We also used the smoothing property from the radiation transfer function, $T(k,\tau) = 3(\sin{l}-l\cos{l})/l^{3}$, where $l=k\tau/\sqrt{3}$ and $\tau=1/aH$, to define the new power spectrum form as $\tilde{\Delta}^{2}_{\zeta}(k) = T^{2}(k,r_{m})\Delta^{2}_{\zeta}(k)$. \textcolor{black}{Such use of the transfer function allows us to implement a damping effect, after going to the Fourier space, over the constantly evolving sub-horizon modes after they re-enter the horizon at scales $l=kc_{s}\tau \gg 1$. The structure of the transfer function tells us that in the sub-Horizon limit, $l\gg 1$, we have $T(k,\tau)=0$, and for scales in the super-Horizon limit, $l\ll 1$, we have $T(k,\tau)=1$ which makes it clear why the role of the transfer function remains valid only inside the Horizon and not on the super-Horizon mode evolution. }
It is important to note that eqns. (\ref{sigcr}-\ref{sigr}) are evaluated at the horizon re-entry scale, $r_{m} = (\tilde{c_{s}}k_{H})^{-1}$, which corresponds to the wavenumber during the formation of PBH and the scale where the compaction function maximizes. \textcolor{black}{Soon as the threshold condition becomes satisfied, at $\tau=r_{m}$ the collapse of perturbations proceeds quickly and thus the modes which finally satisfy, $k\sim {\cal O}(r_{m}^{-1})$, lead to the dominant contributions towards PBH formation \cite{Musco:2018rwt}.} The relation between $\tilde{c_{s}}k_{H} \propto 1/\sqrt{M_{H}}$ \cite{Sasaki:2018dmp,Kawasaki:2016pql} induces $M_{H}$ dependence into the mass fraction, see appendix (\ref{App:B}).

Fig.(\ref{2dpdffig}) depicts contour plots of the 2D-PDF for different horizon mass values. \textcolor{black}{In order to visualize the PDF, we use the eqn. (\ref{PDF}). This formula shows that various correlation functions depend on the power spectrum amplitude, which we fix as $A=10^{-2}$, the minimum value necessary to facilitate PBH production. The overall shape of the PDF depends on the set of such parameters. Although varying the power spectrum amplitude influences the PDF features, the amplitude is not directly fixed by the Gaussian variables $C_{G},\zeta_{G}$. However, the significance of the chosen amplitude comes into play when we set out to determine the domain of integration, ${\cal D} = \{{\cal C}(r) \geq {\cal C}_{th} \wedge {\cal C}_{G}(d\zeta/d\zeta_{G}) \leq 2u\}$, which requires scaling the PDF points with the elements, $\sigma_{c},\sigma_{r}$, in order to solve the inequalities for the Gaussian variables, $C_{G},\zeta_{G}$. The above domain ${\cal D}$ results from maximising the compaction function and on which more details are provided in appendix \ref{App:B}.} The purple, orange, and green coloured lines distinguish the allowed domains of integration for the mass fraction in eqn.(\ref{Beta}). \textcolor{black}{Each domain for a given threshold shares one common dashed black boundary with the respective colour for that threshold forming a complete integration region shaded in cyan.} We observe that as the threshold on the compaction function, ${\cal C}_{\rm th}$, increases, the domain gets more squeezed and moves farther from the center of the contour. The greater the support of the domain within the PDF, the greater the probability of having a sizeable abundance of that particular PBH mass, which is slightly less than the horizon mass at re-entry. The contours also show a notable behavior where the quantity $\gamma_{cr}$ increases as we go below, and also above, the mass, $M_{H} \sim {\cal O}(10^{-3}M_{\odot})$, which signals large correlations between the two Gaussian variables. \textcolor{black}{For masses with, $M_{H} > {\cal O}(0.1 M_{\odot})$, the domain shows no overlap with the PDF, and thus it is much likely that the production of near solar mass PBHs is highly suppressed in our framework.}

    \begin{figure}[htb!]
    	\centering
   {
   \includegraphics[width=9cm,height=7cm] {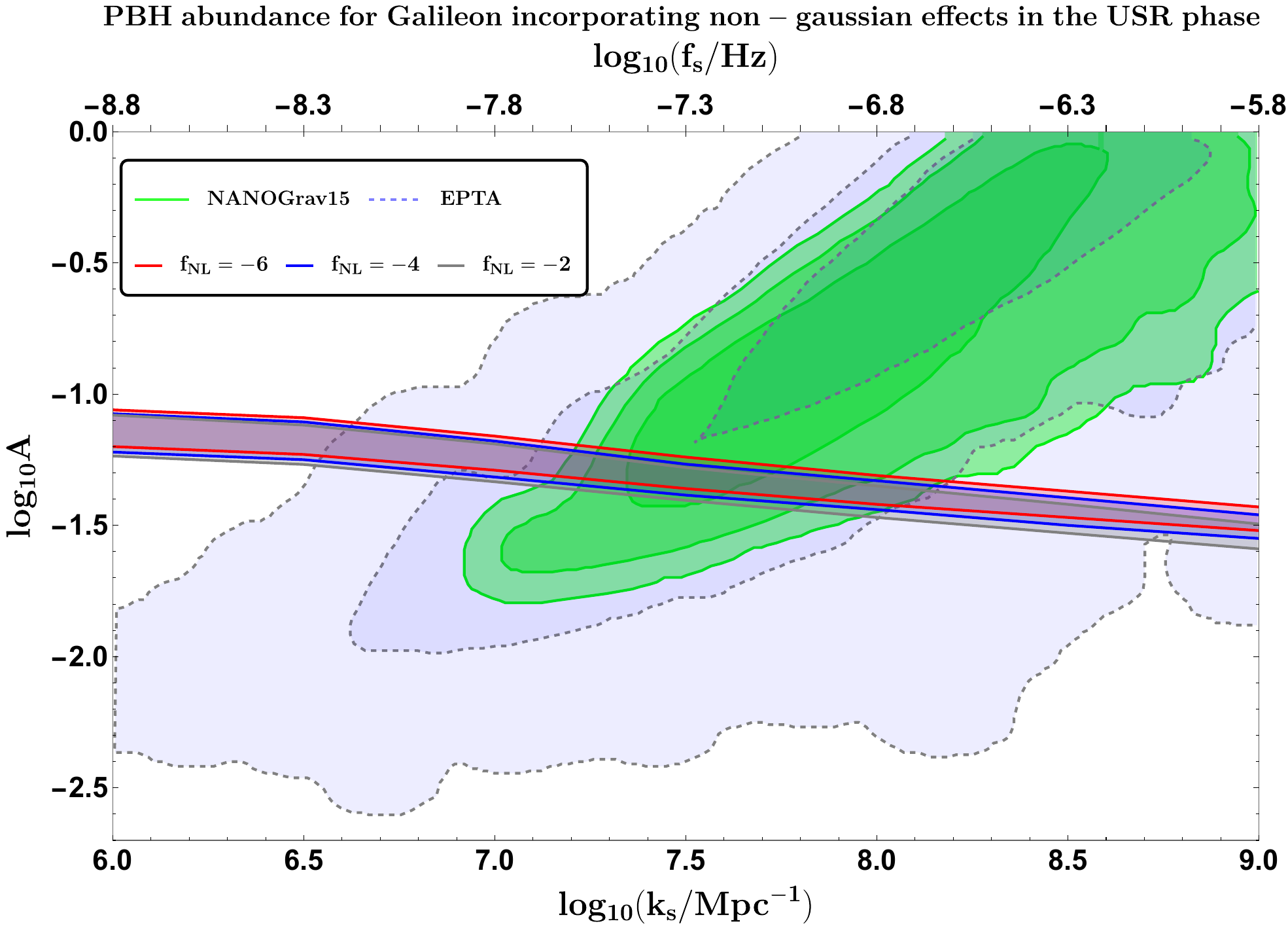}
    }
    	\caption[Optional caption for list of figures]{PBH abundance from Galileon theory for different values of the non-Gaussianity measure, $f_{\rm NL}=-6$ (red), $f_{\rm NL}=-4$ (blue), $f_{\rm NL}=-2$ (gray), and a fixed compaction threshold, ${\cal C}_{\rm th}=0.65$. In each band the PBH abundance belongs to the interval, $f_{\rm PBH} \in (1, 10^{-3})$. The green and light-blue posteriors representing, respectively, the NANOGrav 15 and EPTA.} 
    	\label{pbhfig2}
    \end{figure}

    \begin{figure}[htb!]
    	\centering
   {
   \includegraphics[width=8.5cm,height=7cm] {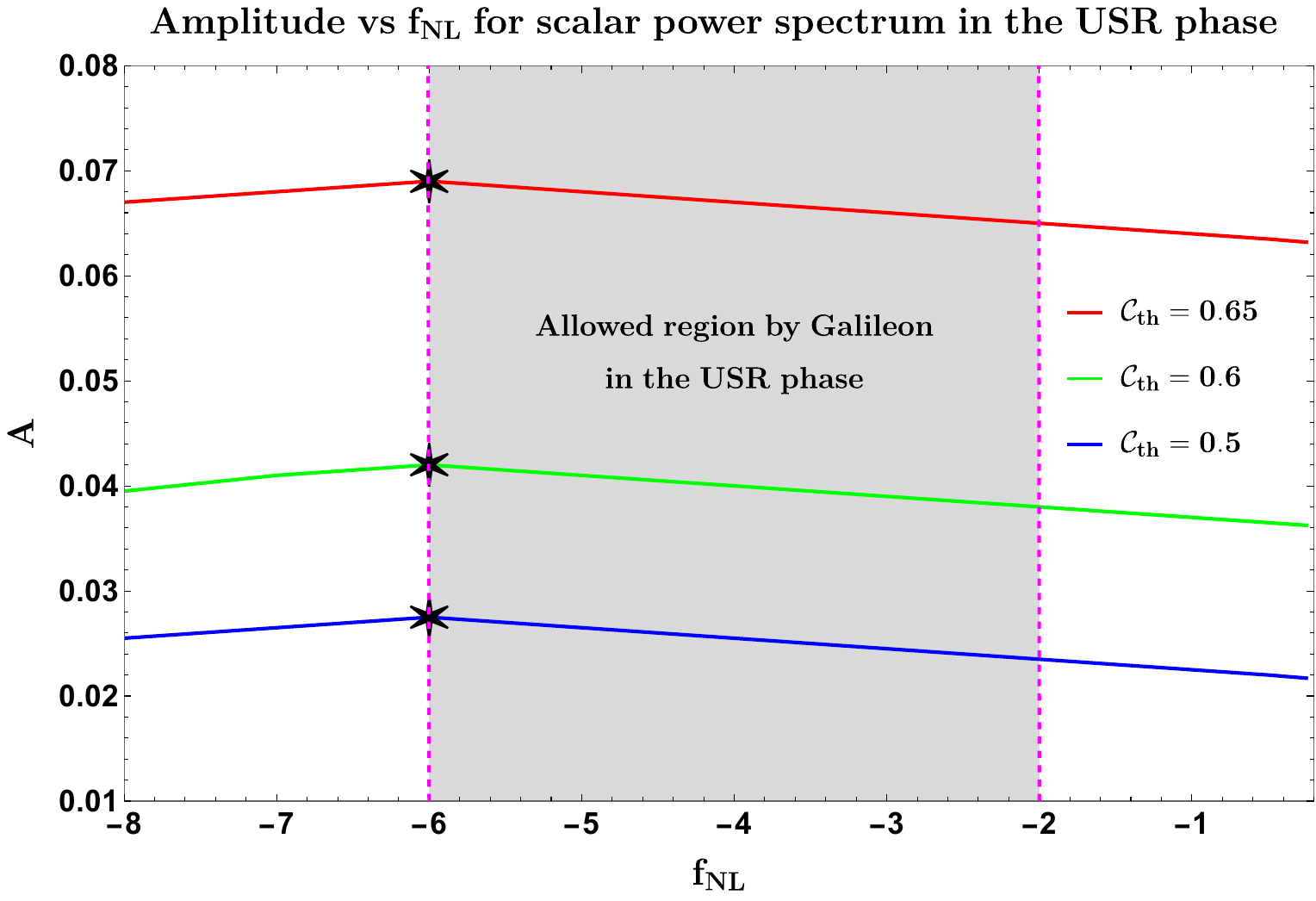}
    }
    	\caption[Optional caption for list of figures]{Behaviour of the peak amplitude in the USR phase with changing values of negative NG, $f_{\rm NL}$. Different values of ${\cal C}_{\rm th}=\{0.65,0.6,0.5\}$ are considered and represented by red, green, and blue lines respectively, while the USR transition wavenumber $k_{s}=10^{7}{\rm Mpc}^{-1}$, and resulting fractional abundance $f_{\rm PBH}=1$ is kept fixed. The black star represents the amplitude value corresponding to the maximum allowed abundance in the present conditions for $f_{\rm NL}=-6$. The gray shaded region with, $-6 \leq f_{\rm NL} \leq -2$, shows the theoretically allowed NG values in Galileon theory during the USR phase.} 
    	\label{ampvsfnl}
    \end{figure}

    \begin{figure}[htb!]
    	\centering
   {
   \includegraphics[width=8.5cm,height=7cm] {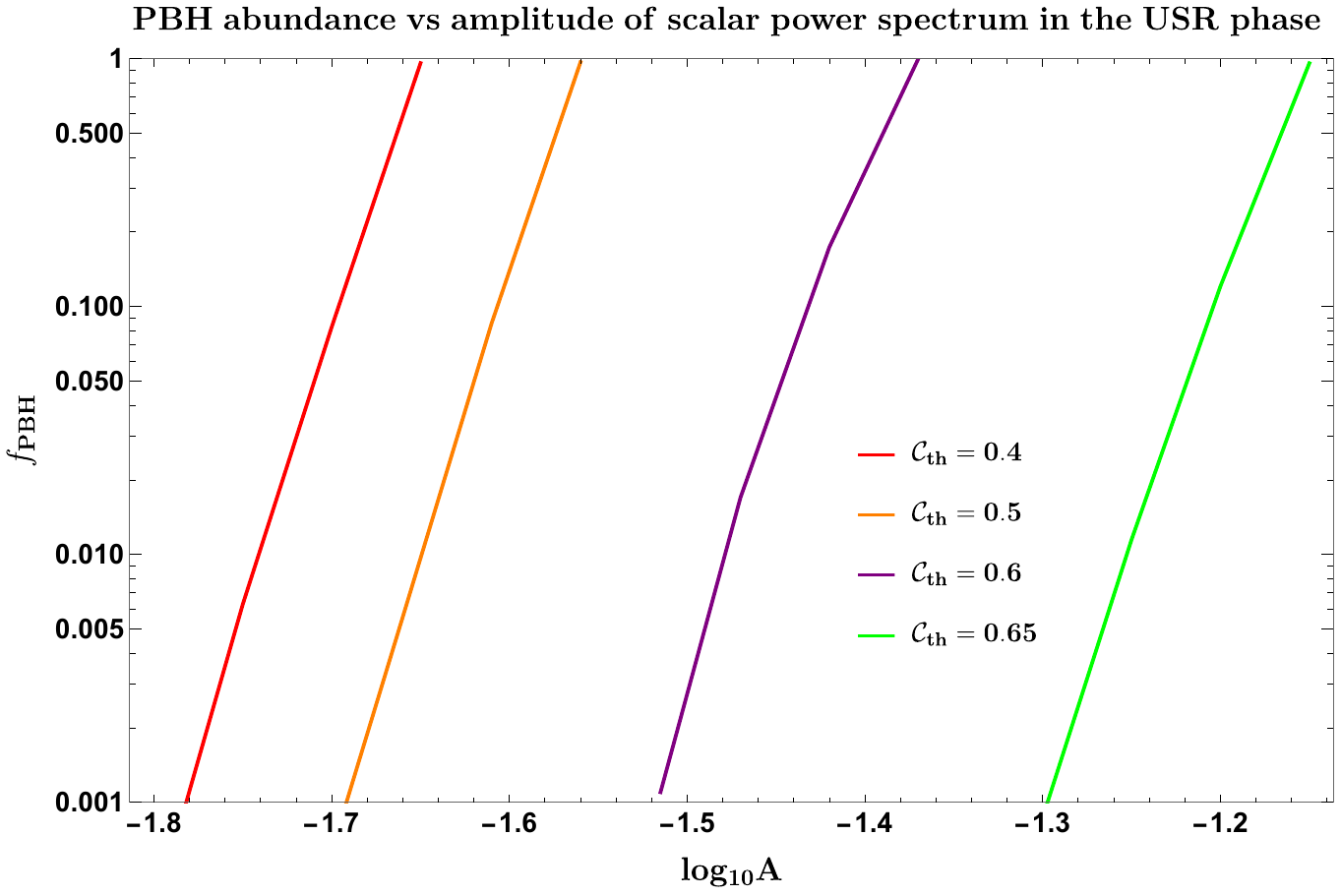}
    }
    	\caption[Optional caption for list of figures]{Dependence of the PBH abundance on the scalar power spectrum amplitude of the curvature perturbation in the USR phase. The red, orange, blue, and green lines correspond to the threshold values of the compaction function, ${\cal C}_{\rm th}=\{0.4,0.5,0.6,0.65\}$, respectively. The transition scale for this plot is fixed at $k_{s}=10^{7}{\rm Mpc}^{-1}$.} 
    	\label{fpbhvamp}
    \end{figure}


\textcolor{black}{Before analyzing the results after computing PBH abundance, we clarify that throughout the remaining analysis, we follow the results in \cite{Musco:2020jjb} on the interval of the threshold values. We maintain the underlying idea of only considering the NGs from the non-linearities in the compaction threshold at Horizon crossing. We elaborate on this with a discussion in the appendix \ref{App:B}.}
  
\textcolor{black}{Estimating the PBH abundance requires knowledge of the mass fraction and its allowed domain. We now analyze our results after using eqn.(\ref{fpbh}) as depicted in fig.(\ref{pbhfig}). The figure shows how large negative NG affects the PBH abundance, related to the transition wavenumbers $k_{s}$, by forcing the amplitude $A$ of the scalar power spectrum in the USR to change for multiple choices of the compaction threshold. For ${\cal C}_{\rm th}=0.65$, we observe that the no overproduction region (brown band and below) gets pushed towards larger amplitudes while maintaining the necessary perturbative approximation for the USR, $\Delta{\cal N}_{\rm USR} \sim {\cal O}(2)$, and where it begins to favour the SIGW interpretation of the NANOGrav15 data within $1\sigma$. Upon changing the choice to ${\cal C}_{\rm th}=0.6$, the region producing sizeable PBH abundance (orange band), $f_{\rm PBH} \in (1, 10^{-3})$, begins to fall in amplitude just outside the $1\sigma$ contour of the NANOGrav15 signal. Till now, we infer from the above that large NG, $f_{\rm NL}\sim {\cal O}(-6)$, tends to lessen the tension for extreme values of the compaction threshold in our theory. As we reach ${\cal C}_{\rm th}=0.5$, PBH production increases further such that the region avoiding overproduction (purple band and below) moves down just close to the $2\sigma$ region. This behaviour is understandable, as increasing the threshold would require more amplitude to generate the same PBH mass. We must note here that the band decreases in amplitude for higher wavenumbers. Since the mass of PBH is related inversely to the transition wavenumber squared, smaller masses require a lesser amplitude of the power spectrum to generate a sizeable abundance.}

\textcolor{black}{We now turn our attention to the results where NG plays a crucial role in determining the power spectrum amplitude and the final fractional abundance of PBH. Fig.(\ref{pbhfig2}) compares the PBH abundance corresponding to distinct values of NG parameter, namely $f_{\rm NL} \in \{-2,-4,-6\}$ for a fixed threshold value of $C_{\rm th}=0.65$. It becomes clear that regardless of the change in $f_{\rm NL}$, there is no major statistical difference in the amplitude of the scalar power spectrum and, consequently, the corresponding PBH abundance. The values between the red lines corresponds to $f_{\rm NL}=-6$, while the values between the blue and gray lines represents $f_{\rm NL}=-4$, and $f_{\rm NL}=-2$ respectively. All such bands lie within the $1\sigma$ contour of the NANOGrav15 and $2\sigma$ contour of the EPTA. The upper red lines lie closer to the $1\sigma$ contour than the blue and gray lines, indicating that the largest scalar power spectrum that can be obtained to get ideal abundance $(f_{\rm PBH}=1)$  is for $f_{\rm NL}=-6$. Upon analyzing, We will see a similar trend for other values of $C_{\rm th}$, and this feature remains consistent with the results in fig.(\ref{ampvsfnl}). Thus, due to the value of $f_{\rm NL}$ only allowed to vary inside a finite interval for Galileon inflation, $-6 \leq f_{\rm NL}\leq -2$, leads to only a small change in the necessary peak amplitude $A$. Also, in the absence of any primordial NGs $(f_{\rm NL}=0)$ and just considering the impact of non-linearities, the SIGW interpretation of the NANOGrav15 signal breaks, and we will not obtain the desirable amplitude of the power spectrum as strictly proven in \cite{Franciolini:2023pbf}.} 

\textcolor{black}{We next elaborate on the fig.(\ref{ampvsfnl}) that explores the impact of NGs on PBH abundance in a different manner.} This plot depicts how the peak amplitude of the power spectrum, in the USR, changes for different values of local NG to achieve a maximum possible abundance. Large negative NG can seed large fluctuations that generate higher-mass black holes. Here, we have focused on the masses of PBH formed with the transition scale fixed at $k_{s}=10^{7}{\rm Mpc}^{-1}$. This scale allows for the generation of $M_{\rm PBH} \sim {\cal O}(10^{-2}M_{\odot})$. As we tend to decrease $f_{\rm NL}$ below $-6$, we observe that the amplitude in the USR falls to saturate the abundance of the said PBH masses. On the contrary, raising $f_{\rm NL}$ above $-6$ will increase the abundance of masses in the lower end of the allowed PBH mass spectrum, which then requires less amplitude for their formation. In the present context, Galileon theory does not allow for NG values outside the interval, $-6 \leq f_{\rm NL} \leq -2$, and the gray shaded region represents the theoretically allowed region. Throughout the analysis in fig.(\ref{pbhfig}), we have considered taking $f_{\rm NL}=-6$, its related power spectrum amplitudes shown using a black star marker for various thresholds, ${\cal C}_{\rm th} = \{0.5,0.6,0.65\}$, and this has provided us with the most satisfactory results for the agreement of the generated SIGWs with the PTA data. 

\textcolor{black}{After discussing the features of PBH abundance for various transition wavenumbers, including the impact of changing primordial NGs, we look at the sensitivity of $f_{\rm PBH}$ to the amplitude $A$ more closely.} In the fig.(\ref{fpbhvamp}) above, we can see how the amplitude changes for particular values of the compaction threshold to give rise to the fractional abundance in between the interval $f_{\rm PBH} \in (10^{-3},1)$, when we have fixed the transition scale $k_{s}=10^{7}{\rm Mpc}^{-1}$. As one can anticipate, increasing the threshold requires larger amplitudes of the scalar power spectrum to seed sizeable abundances of black holes. The plot also demonstrates that the abundance is highly sensitive to the amplitude, changing quickly and saturating to unity in an extremely short interval of the amplitude values. Here, we have kept the analysis within the mentioned regime of threshold values, ${\cal C}_{\rm th}$, where cosmological perturbation theory holds good.

From our examination of the abundance of PBH, we observe that unlike working with a two-field model, like the curvaton, for generating the PBH forming curvature perturbations, the single-field Galileon theory employs an additional USR feature during inflation, whose duration respects the established perturbativity conditions \cite{Choudhury:2023hvf,Choudhury:2023kdb,Choudhury:2023hfm}. It sufficiently enhances the scalar power spectrum for PBH formation and contributes towards the overall quantum one-loop corrections. The resulting amplitude corresponds to sizeable PBH abundance without risking overproduction.
The analysis done for the curvaton model in \cite{Franciolini:2023pbf, Ferrante:2022mui} considers the NGs in great detail to obtain a suitable abundance and thus leads to the amplitude of the power spectrum that is enough to alleviate the tension between NANOGrav 15 and SIGWs. On the other hand, Galileon further includes larger negative NGs such that the resulting amplitude generates the SIGWs that correlate strongly with NANOGrav 15 data in the presence of the quantum loop effects.

In this letter, we have utilized the threshold statistics on the compaction function, which helps us to consider the primordial NGs and the inbuilt non-linearities into the density contrast field.
We first demonstrated that the SIGW spectra corresponding to $c_s = \{0.04,0.05,0.06\}$ comply well with the NANOGrav 15 and EPTA signal. Accompanying this, with the NG value of $f_{\rm NL} \sim {\cal O}(-6)$, we obtained sizeable PBH abundance in the range $f_{\rm PBH}\in(10^{-3}-1)$ which manages to remove the overproduction problem. We conclude that large negative NGs within Galileon theory, which come from respecting the perturbativity conditions on the USR phase, allow for a sizeable PBH abundance and consequently lead to the amplitude of scalar power spectrum that conforms within $2\sigma$ to the SIGW interpretation of the observed GW signal from the PTA collaborations.

{\bf Acknowledgements:} SC would like to thank The National Academy of Sciences (NASI), Prayagraj, India for being elected as a member of the academy. SP is supported by the INSA Senior scientist position at NISER, Bhubaneswar through the Grant number INSA/SP/SS/2023. MS is supported by Science and Engineering Research Board (SERB), DST, Government of India under
the Grant Agreement number CRG/2022/004120 (Core Research Grant). MS is also partially supported by the
Ministry of Education and Science of the Republic of Kazakhstan, Grant No. 0118RK00935, and CAS President’s
International Fellowship Initiative (PIFI). 

\section*{Appendix}

\appendix


\section{Details for the scalar power spectrum}\label{App:A}

\textcolor{black}{In this section we focus on building the comoving curvature perturbation mode solutions once we obtain the corresponding equation of motion for the curvature perturbation from the second-order action presented in eqn.(\ref{quadaction}).} Redefining the action in terms of a new variable, $z$, helps us to give the resulting equation of motion:
\bea
&&\zeta^{''}_{\bf k}(\tau)+2\frac{z^{'}(\tau)}{z(\tau)}\zeta^{'}_{\bf k}(\tau) +c_{s}^{2}k^{2}\zeta_{\bf k}(\tau)=0, \nonumber\\
&& \text{where}\quad z(\tau) = a\sqrt{2{\cal A}}/H^{2},
\eea
with this equation in hand, we can proceed towards analyzing the mode solutions and the related power spectrum functions in different phases during inflation in our theory.

The power spectrum receives contributions from each SRI, USR, and SRII phase. However, we encounter a different set of Bogoliubov coefficients for each phase after solving the Israel junction conditions for the curvature perturbation modes. These new coefficients correctly describe the mode solutions of the new phase, including the properties of the respective quantum vacuum state. The mode solutions, along with their corresponding Bogoliubov coefficients for each phase, are listed below. We introduce the following shorthand notations for the sake of convenience while writing the mentioned mode solutions and coefficients:
\bea
f_{\pm,\bf k}(\tau) &=& \left(\frac{iH^{2}}{2\sqrt{\cal A}}\right)\frac{1}{(c_{s}k)^{3/2}}\times(1\pm ikc_{s}\tau)\exp{(\mp ikc_{s}\tau)}, \nonumber\\
&& \quad {\cal X}_{s} \equiv k/k_{s},\quad\quad {\cal X}_{e} \equiv k/k_{e},
\eea
where, $|{\bf k}| = k$, denotes the magnitude of the wavenumber.
\begin{enumerate}

\item \underline{\textbf{For the SRI phase:}}
    The following expression represents the solution for the scalar mode corresponding to the SRI phase where the conformal time satisfies the condition, $\tau < \tau_{s}$: 
    \bea  \zeta_{\bf {k}}(\tau) = \left[\alpha^{(1)}_{\bf {k}}f_{+,\bf k}(\tau) - \beta^{(1)}_{\bf {k}}f_{-,\bf k}(\tau)\right]. \eea
    The Bogoliubov coefficients involved in the above solution have the following values:
     \bea \label{s21a}\alpha^{(1)}_{\bf k}=1,
        \quad\quad\quad  \label{s21b}\beta^{(1)}_{\bf k}=0, \eea
    \textcolor{black}{This choice of coefficients indicate use of the well-studied Bunch-Davies quantum vacuum state as our starting point. }
\item \underline{\textbf{For the USR phase:}}
    The following expression represents the solution for the scalar mode corresponding to the USR phase where the conformal time satisfies the condition, $\tau_{s} \leq \tau < \tau_{e}$:
    \bea   \zeta_{\bf {k}}(\tau)&=&\left(\frac{\tau_{s}}{\tau}\right)^{3} \left[\alpha^{(2)}_{\bf k}f_{+,\bf k}(\tau) - \beta^{(2)}_{\bf k}f_{-,\bf k}(\tau) \right].
    \eea    
    The Bogoliubov coefficients involved in the above solution have the following values:
    \bea
        \label{s22a}\alpha^{(2)}_{\bf k}&=& 1+\frac{3}{2 i {\cal X}_{s}^{3}}\left(1+{\cal X}_{s}^{2}\right),\nonumber\\
        \label{s22b}\beta^{(2)}_{\bf k}&=& \frac{3}{2 i {\cal X}_{s}^{3}}\left(1-i{\cal X}_{s}^{2}\right)^{2}\exp{\left(2i{\cal X}_{s}\right)}. \eea
    \textcolor{black}{To obtain the above we require applying the continuity of the mode solutions and its conjugate momenta in SRI and USR at the transition scale $k_{s}$.}

\item \underline{\textbf{For the SRII phase:}}
    The following expression represents the solution for the scalar mode corresponding to the SRII phase where the conformal time satisfies the condition, $\tau_{e} \leq \tau \leq \tau_{\rm end}$: 
    \bea  \zeta_{\bf {k}}(\tau)&=&\left(\frac{\tau_{s}}{\tau_{e}}\right)^{3} \left[\alpha^{(3)}_{\bf k}f_{+,\bf k}(\tau) - \beta^{(3)}_{\bf k}f_{-,\bf k}(\tau) \right]. \eea
    The Bogoliubov coefficients involved in the above solution have the following values:
    \bea
        \alpha^{(3)}_{\bf k}&=&-\frac{1}{4{\cal X}_{s}^{3}{\cal X}_{e}^{3}}\Bigg[9 \left(-{\cal X}_{e}+i\right){}^2 \left({\cal X}_{s}+i\right){}^2 \exp{\left(2 i 
        \left({\cal X}_{s}-{\cal X}_{e}\right)\right)}\nonumber\\
        &+&
        \left\{{\cal X}_{e}^2\left(2{\cal X}_{e}+3i\right)+3i\right\}\left\{{\cal X}_{s}^2\left(-2{\cal X}_{s}+3i\right)+3i\right\}\Bigg],\\
        \label{s23b}\beta^{(3)}_{\bf k}&=&\frac{3}{4{\cal X}_{s}^{3}{\cal X}_{e}^{3}}\Bigg[\left({\cal X}_{s}+i\right)^2\left\{{\cal X}_{e}^{2}\left(2i{\cal X}_{e}+3\right)\right\}\exp{\left(2i{\cal X}_{s}\right)}\nonumber\\
        &+&i\left({\cal X}_{e}+i\right)^2\left\{3i+{\cal X}_{s}^{2}\left(-2{\cal X}_{s}+3i\right)\right\}\exp{\left(2i{\cal X}_{e}\right)}\Bigg].\quad\quad
    \eea
    \textcolor{black}{In order to obtain the above we require applying the continuity of the mode solutions and its conjugate momenta in USR and SRII at the transition scale $k_{e}$.}
\end{enumerate}

From using the above information, one can able to write the final scalar power spectrum which includes the tree-level and the one-loop contributions in terms of the wavenumber as:
\begin{widetext}
    \bea \Delta^{2}_{\zeta,{\bf Total}}(k) &=& \left[\Delta^{2}_{\zeta,{\bf Total}}(k)\right]_{\bf SRI} + \left[\Delta^{2}_{\zeta,{\bf Total}}(k)\right]_{\bf USR}\Theta(k-k_{s}) +\left[\Delta^{2}_{\zeta,{\bf Total}}(k)\right]_{\bf SRII}\Theta(k-k_{e}),\nonumber\\
&=& \left[\Delta^{2}_{\zeta,{\bf Tree}}(k)\right]_{\bf SRI}\bigg\{1+ \left(\frac{k_{e}}{k_{s}}\right)^{6}\bigg(\big|\alpha_{\bf k}^{(2)}-\beta_{\bf k}^{(2)}\big|^2\Theta(k-k_s)+\big|\alpha_{\bf k}^{(3)}-\beta_{\bf k}^{(3)}\big|^2 \;\Theta(k-k_e)\bigg) + {\cal Q}_{c}\bigg\},\nonumber\\
&=& A\left[1 + {\cal X}_{s}^{2}\right]\bigg\{\left(\frac{k_{s}}{k_{e}}\right)^{6}(1+{\cal Q}_{c}) + \bigg(\big|\alpha_{\bf k}^{(2)}-\beta_{\bf k}^{(2)}\big|^2\Theta(k-k_s)+\big|\alpha_{\bf k}^{(3)}-\beta_{\bf k}^{(3)}\big|^2 \;\Theta(k-k_e)\bigg)\bigg\},
\eea
\end{widetext}
where we have used the following definitions:
\bea
\left[\Delta^{2}_{\zeta,{\bf Tree}}(k)\right]_{\bf SRI} &=& \left(\frac{H^{4}}{8\pi^{2}{\cal A}c_{s}^{3}}\right)_{*}\left[1+{\cal X}_{s}^{2}\right], \nonumber\\
A &=& \left(\frac{H^{4}}{8\pi^{2}{\cal A}c_{s}^{3}}\right)_{*}\left(\frac{k_{e}}{k_{s}}\right)^{6}, \eea
and the term labeled ${\cal Q}_{c}$ represents the collection of the one-loop quantum corrections from each phase. For more details on the explicit computation of these one-loop contributions in Galileon theory, see ref.\cite{Choudhury:2023hvf}. \textcolor{black}{Also, aside from the Bogoliubov coefficients for each phase which includes the sharp transition wavenumbers, the total scalar power spectrum from Galileon theory takes in the effective sound speed $c_{s}$, and the time-dependent coefficients obtained before for the second-order action in eqns.(\ref{coeffA},\ref{coeffB}).}

\section{Compaction function and its threshold statistics}\label{App:B}
This section details the necessary computations when applying the threshold statistics to the compaction function. First introduced by Shibata and Sasaki in \cite{Shibata:1999zs}, this function is defined to be twice the excess mass over the areal radius:
\bea \label{compaction}
{\cal C}(r,t)&=& 2 \times\bigg(\frac{M(r,t)-M_{o}(r,t)}{R(r,t)} \bigg)\nonumber\\
&=&\frac{2}{R(r,t)}\int_{S_{R}^{2}} d^{3}\Vec{x} \rho_o(t)\delta(\Vec{x},t),
\eea
where $M_{o}(r,t)=\rho_{0}(t)V_{o}(r,t)$, with $V_{o}(r,t)=(4\pi/3)R(r,t)^{3}$, denotes the background mass with respect to the energy density $\rho_{o}(t)$, and $M(r,t)$ denotes the Misner-Sharp mass within the sphere $S_{R}^{2}$ of areal radius, $R(r,t)\equiv a(t)r\exp{(\zeta(r))}$, written in the comoving radial coordinates. 
Now, upon analyzing the perturbations well outside the cosmological horizon, one can accommodate the non-linearities from the conserved curvature perturbation, $\zeta(r)$, into the density contrast field $\delta(\vec{x},t) = \delta\rho(\vec{x},t)/\rho_{o}(t)$ by using the long-wavelength approximation (also known as the gradient expansion) on such super-horizon scales. For a generic background with the equation of state $w$, the above-mentioned non-linearities in the radial coordinates result from the expression \cite{Musco:2018rwt,Harada:2015yda}:
\bea \label{NLdensity}
\delta(r,t) &=& -\frac{2u}{3}\left(\frac{1}{aH}\right)^{2}e^{-5\zeta(r)/2}\nabla^{2}e^{\zeta(r)/2},\nonumber\\
&\approx& -\frac{2u}{3}\frac{1}{(aH)^2}e^{-2\zeta(r)} \bigg[\zeta^{''}(r) + \frac{2}{r}\zeta^{'}(r) + \frac{1}{2}\zeta^{'}(r)^{2}\bigg],\quad\quad
\eea
where $u=3(1+w)/5+3w$ and a prime indicates a radial derivative. The analysis performed in this letter is not restricted to any particular component but can be generalized to any equation of state, along with Galileon, within $0 \leq w \leq 1$, which covers $w = \{0,1/3,1\}$ describing the matter-dominated,  radiation dominated, and the kination era respectively. In the above equation, the Laplacian in radial coordinates and the exponential terms introduce the explicit non-linear effects, which makes it clear that NGs will inevitably be present regardless of the statistics of the curvature perturbation $\zeta(r)$ \cite{Young:2019yug, DeLuca:2019qsy}. Using the non-linear eqn.(\ref{NLdensity}), the compaction function in eqn.(\ref{compaction}) achieves the new time-independent definition: 
\bea \label{compaction2}
{\cal C}(r)=-2ur\zeta'(r)\bigg(1+\frac{r}{2}\zeta'(r)\bigg).
\eea
The above equation can be separated into its linear and non-linear components as follows:
\bea
{\cal C}(r)&=&{\cal C}_{\rm L}(r)-{\cal C}_{\rm NL}(r)\quad{\rm where},\nonumber\\
{\cal C}_{\rm L}(r)&=&-2ur\zeta'(r),\quad{\rm and}\quad {\cal C}_{\rm NL}(r)={\cal C}_{\rm L}(r)^2/4u.
\eea
Now, the criterion for the PBH formation comes from maximizing the compaction function at the scale $r_{m}$, which, on using the eqn.(\ref{compaction2}), enables the condition: 
\bea \label{maxcond}
\zeta'(r_{m})+r_{m}\zeta''(r_{m})=0.
\eea
The same comoving scale $r_{m}$ can as well be associated to the Horizon mass $M_{H}$ through the standard relation in terms of the corresponding wavenumbers, $\tilde{c_{s}}k_{H} = r_{m}^{-1}$ , that source the PBH forming perturbation during horizon re-entry (crossing):
\bea
M_H \approx 17M_{\odot}\bigg[\frac{g_{*}}{10.75}\bigg]^{-1/6}\bigg[\frac{\tilde{c_{s}}k_{H}}{10^6 \rm Mpc^{-1}}\bigg]^{-2}.
\eea
where $g^{*}=106.75$ represents the relativistic d.o.f in the radiation-dominated epoch, and, in the present scenario, $k_{H}$ corresponds to the transition scale $k_{s}$ into the USR region. We incorporate the effective sound speed at the transition scale by using $\tilde{c_{s}} \approx 1\pm\delta$, such that $\delta \ll 1$. 
At this stage, let's also examine the excess mass volume-averaged over a region with areal radius $R$. This quantity has a special relation at the scale $r_{m}$, and the corresponding re-entry time $t_{H}$, with the compaction function at the same scale:
\bea \label{reentryeql} 
\delta_{m} &=& \frac{\delta M(r,t)}{M_{o}(r,t)}
\equiv \frac{3}{R_{m}}\int^{R_{m}}_{0}\frac{\delta\rho}{\rho_{o}}R^{2}dR \nonumber\\
&=& {\cal C}(r_{m}) = 3\delta(r_{m},t_{H}),
\eea
in the above, $R_{m} = a(t_{H})r_{m}e^{\zeta(r_{m})}$, is the physical length scale of the perturbation during re-entry. The last two equalities are realised only after the use of the Horizon crossing condition $R_{m}H(t_{H})=1$, the eqn.(\ref{NLdensity}), and the condition in eqn.(\ref{maxcond}) in above while considering eqn.(\ref{compaction2}).
The eqn.(\ref{reentryeql}) is a significant result, as it directly brings the volume-averaged density contrast, $\delta_{m}$, into the picture through the compaction function at horizon re-entry.

Current estimates for the volume-averaged density contrast on superhorizon scales, obtained using numerical analysis, result in the interval, $2/5 \leq \delta_{\rm th} \leq 2/3$, where under the first approximation, that is neglecting higher-order effects in the gradient expansion approach, is said to provide a reasonable outcome for the perturbation amplitude during horizon re-entry \cite{Musco:2018rwt, Musco:2020jjb} at the associated scale $r_{m}$ and the perturbation mode $k_{H}$. An important fact to remember here is that in the mentioned studies, only the NG effects due to the non-linearities from eqn.(\ref{NLdensity}) are taken into account, along with assuming that the curvature perturbation only follows Gaussian statistics to give us the above interval for $\delta_{\rm th}$. Recent studies in \cite{Escriva:2022pnz,Kehagias:2019eil} have looked into the effects of quadratic primordial NGs on the computation of the threshold and found changes only at a few percent level. Through the equality in eqn.(\ref{reentryeql}), we can then directly determine the range of the threshold values ${\cal C}_{\rm th}$.

\textcolor{black}{We now discuss briefly the approaches adopted by different studies to incorporate the impact of $f_{\rm NL}$ on the compaction threshold estimation. The study in \cite{Escriva:2022pnz} makes it clear, through the use of a curvature perturbation profile, that the volume-averaged compaction function approach with its universal threshold value of $\bar{C}=2/5$, only works well for models with positive $f_{\rm NL}$ and fails for the cases where negative $f_{\rm NL}\lesssim -1$. To thoroughly determine the compaction threshold for a specific model would require using a curvature perturbation profile, $\zeta_{G}$, and dedicated simulations on the behavior of the compaction function, which shows the resulting impact coming from the presence of quadratic NGs. The curvature perturbations will also then experience changes with the new threshold values for different values of $f_{\rm NL}$ and go on to directly affect the PBH formation process.  Otherwise, one can also follow the relatively new path-integral techniques followed by \cite{Kehagias:2019eil} where they employ threshold statistics to study non-Gaussian corrections to the curvature perturbation correlations and their impact on the average density perturbation profile. For this work, we avoid this entirely separate study of estimating the threshold and follow the volume-averaged compaction threshold prescription as laid in \cite{Musco:2020jjb}. For our analysis, we do not begin by assuming a template expression for the Gaussian curvature perturbation and rely on the various correlations and their joint PDF to get the set of values for the variables, $C_{G},\zeta_{G}$. We then use the horizon crossing property to identify the two quantities, $\delta_{m}={\cal C}(r_m)$, and get the threshold ${\cal C}_{\rm th}$ to ultimately perform the integration over the complete domain satisfying specific conditions which we can now begin to examine. } 

Since we are considering a generic expression for the curvature perturbation $\zeta(r)$ which carries within the primordial NG information, this also appears inside the equation (\ref{compaction2}), which we now mention as: 
\bea
{\cal C}(r) = {\cal C}_{G}(r)\frac{d\zeta}{d\zeta_{G}} - \frac{1}{4u}\left({\cal C}_{G}(r)\frac{d\zeta}{d\zeta_{G}}\right)^{2},
\eea 
with ${\cal C}_{G}(r)= -2ur\zeta'_{G}(r)$.
The conditions under which a cosmological perturbation undergoes collapse to form PBH get decided by, firstly, requiring the condition ${\cal C} \geq {\cal C}_{\rm th}$, based on the threshold of the compaction function ${\cal C}_{\rm th}$. 
While this is true, maximizing the compaction function sets another boundary condition to provide us with a proper domain of values:  
\bea
{\cal C}'(r)&=& {\cal C}_{\rm L}'(r)-{\cal C}_{\rm NL}'(r)=
 {\cal C}_{\rm L}'(r)-\frac{d}{dr}\bigg(\frac{{\cal C}_{\rm L}(r)^2}{4u}\bigg),\nonumber\\
 &=&
 {\cal C}_{\rm L}'(r) - \frac{1}{2u}\bigg({\cal C}_{\rm L}(r)\;{\cal C}_{\rm L}'(r)\bigg)=0,\nonumber\\
 \implies {\cal C}_{\rm L}(r) &=& 2u.
\eea
Finding a double derivative gives an affirmative over the second boundary condition as being, ${\cal C}_{\rm L}(r) \le 2u$ or $ {\cal C}_{G}(d\zeta/d\zeta_{G}) \leq 2u$.
As a result, the complete domain for integrating the PBH mass fraction after the use of threshold statistics on the compaction function reads: ${\cal D} = \{{\cal C}(r) \geq {\cal C}_{th} \wedge {\cal C}_{G}(d\zeta/d\zeta_{G}) \leq 2u\}$. 

\section{Primordial Non-Gaussianity and PBH Mass fraction}\label{App:C}

Including NG in the comoving curvature perturbation is essential due to the presence of a USR phase during inflation. Now, the NG corrections are suppressed at the CMB scales due to the small values of the slow-roll parameters. However, the introduction of a USR phase contributes significantly to the NGs. In this section, we elaborate on the effects of deviating from the Gaussian limit due to considering the local type of primordial NGs in the curvature perturbations. Specifically, we are considering the quadratic parameterization of the perturbation, $\zeta = \zeta_{G} + (3/5) {f_{\rm NL}}\zeta_{G}^{2}$, in terms of the Gaussian random variable $\zeta_{G}$.  

Starting with the already discussed 2D joint PDF in eqn.(\ref{PDF}), we ask how deviating from a perfectly correlated scenario, where one essentially considers a Dirac delta type of distribution, brings changes into the results of the primordial abundance of black holes. Upon Taylor expanding the existing PDF in between the $2$ Gaussian variables, ${\cal C}_{G},\zeta_{G}$, around the value of $\gamma_{cr}=1$, we can get the following expression:
\begin{widetext}
\bea \label{taylorPg}
\mathds{P}_{G}({\cal C}_{G},\zeta_{G}) &=& \frac{1}{2 \sqrt{2} \pi \sigma_{c}\sigma_{r} \sqrt{1-\gamma_{cr}}}\exp \bigg(\displaystyle{\frac{1}{4 (\gamma_{cr}
   -1)}\left(\frac{{\cal C}_{G}}{\sigma_{c}}-\frac{\zeta_{G}}{\sigma_{r}}\right)^{2}+\sum_{n=3}^{\infty}\left(\frac{-1}{2}\right)^{n}\left(\frac{{\cal C}_{G}}{\sigma_{c}}+\frac{\zeta_{G}}{\sigma_{r}}\right)^{2}(\gamma_{cr}-1)^{n-3}}\bigg)\nonumber\\
   && \times\left(\displaystyle{1+\frac{1-\gamma_{cr}}{4}+\frac{3 (1-\gamma_{cr})^{2}}{32} +\frac{5 (1-\gamma_{cr})^{3}}{128}+\frac{35 (1-\gamma_{cr})^{4}}{2048}+\frac{63 (1-\gamma_{cr})^{5}}{8192}+\frac{231 (1-\gamma_{cr})^{6}}{65536}+\cdots}\right), \quad\quad\eea
\end{widetext}
where the $\cdots$ represent the higher-order terms in the Taylor expansion and, under the particular limit $\gamma_{cr} \rightarrow 1$, the above distribution blackuces to as follows:
\bea \label{peakpdf}
&& \lim_{\gamma_{cr}\rightarrow 1}\mathds{P}_{G}({\cal C}_{G},\zeta_{G}) = \frac{1}{\sqrt{2\pi}\sigma_{c}\sigma_{r}}\exp{\left(-\frac{1}{8}\left(\frac{{\cal C}_{G}}{\sigma_{c}}+\frac{\zeta_{G}}{\sigma_{r}}\right)^{2}\right)}\nonumber\\
&&\times \lim_{\gamma_{cr}\rightarrow 1}\frac{1}{\sqrt{2\pi(2(1-\gamma_{cr}))}}\exp{\left(\frac{1}{4(\gamma_{cr}-1)}\left(\frac{{\cal C}_{G}}{\sigma_{c}}-\frac{\zeta_{G}}{\sigma_{r}}\right)^{2}\right)},\nonumber \eea\bea
&&=\frac{1}{\sqrt{2\pi}\sigma_{c}\sigma_{r}}\exp{\left(-\frac{1}{8}\left(\frac{{\cal C}_{G}}{\sigma_{c}}+\frac{\zeta_{G}}{\sigma_{r}}\right)^{2}\right)}\delta\left(\frac{{\cal C}_{G}}{\sigma_{c}}-\frac{\zeta_{G}}{\sigma_{r}}\right)\quad.
\eea
Here the Dirac delta function enforces the condition, ${\cal C}_{G}/\sigma_{c}=\zeta_{G}/\sigma_{r}$, in the remaining quadratic expression blackucing it down to a sharp or highly correlated distribution function, which must be the case since the correlation coefficient $\gamma_{cr}$ reaches to its maximum value of $\gamma_{cr}=1$. The expansion in eqn.(\ref{taylorPg}) is only valid when the distribution is in the vicinity of being highly correlated and is infinitely differentiable, which seems to be the case here, or else the series expansion breaks for lower values of $\gamma_{cr}$. In the present context, the fig.(\ref{2dpdffig}) featuring the contour plots have the value of the correlation coefficient satisfying $0.84 \lesssim \gamma_{cr} < 1$ for $M_{H} \geq 10^{-3}M_{\odot}$ and thus justifies performing such an expansion. The resulting additional terms will contribute to the overall PBH mass fraction and the corresponding fractional abundance, in addition to the contribution from the sharply peaked distribution. 

Given the peaked distribution in eqn.(\ref{peakpdf}), its contribution to the total mass fraction can be written as: 
\bea
\beta_{\rm peak}(M_{H}) &=& \frac{1}{\sqrt{2\pi}\sigma_{c}\sigma_{r}}\int_{\cal D}d{\cal C}_G\;d\zeta_{G}\;{\cal K}({\cal C}-{\cal C}_{\rm th})^{\gamma}\nonumber\\
&\times& \exp{\left(-\frac{1}{8}\left(\frac{{\cal C}_{G}}{\sigma_{c}}+\frac{\zeta_{G}}{\sigma_{r}}\right)^{2}\right)}\delta\left(\frac{{\cal C}_{G}}{\sigma_{c}}-\frac{\zeta_{G}}{\sigma_{r}}\right),\nonumber\\
&=& \frac{1}{\sqrt{2\pi}\sigma_{r}}\int_{\cal D}{\cal K}\left\{\left[g(\zeta_{G})-\frac{1}{4u}g(\zeta_{G})^{2}\right]-{\cal C}_{\rm th}\right\}^{\gamma}\nonumber\\
&\times& \exp{\left(-\frac{\zeta_{G}^{2}}{2\sigma_{r}^{2}}\right)}d\zeta_{G}.
\eea
Due to the Dirac delta function, one obtains an integral over the variable $\zeta_{G}$ where we define the function $g(\zeta_{G})=\displaystyle{\frac{\sigma_{c}}{\sigma_{r}}\zeta_{G}\frac{d\zeta}{d\zeta_{G}}}$ for convenience. The remaining constants, $\{{\cal K}, u, {\cal C}_{\rm th}, \gamma\}$, are the same as defined previously during the first appearance of the joint PDF. This PBH mass fraction further enables the computation of the related fractional abundance after integration over a range of horizon masses, which label here the possible epochs of black hole formation.
The abundance corresponding to the peak distribution is derived through the following expression:
\bea
f^{\rm peak}_{\rm PBH} &\equiv& \frac{\Omega^{\rm peak}_{\rm PBH}}{\Omega_{\rm DM}}=\frac{1}{\Omega_{\rm DM}}\int d\ln{M_{H}}\left(\frac{M_{H}}{M_{\odot}}\right)^{-1/2}\nonumber\\
&\times&\left(\frac{g_{*}}{106.75}\right)^{3/4}\left(\frac{g_{*s}}{106.75}\right)^{-1}\left(\frac{\beta_{\rm peak}(M_{H})}{7.9 \times 10^{-10}}\right).\quad
\eea
In the above the time-dependent horizon mass contained within the Universe of radius $1/H(t)$ at cosmic time $t$ is defined as $M_{H} \equiv M_{H}(t)= (4\pi/3)\rho(t)/H^{3}(t)$, with $\rho(t)$ being the total energy density of the Universe at the same time and $\Omega_{\rm DM}\simeq 0.264$ corresponds to the dark matter density of the Universe. The Gaussian PDF in the above calculation is responsible for getting the maximum possible contribution to the total abundance. The rest of the terms providing deviation from Gaussian nature in the PDF expansion provide smaller contributions to the total mass fraction and consequently result in an estimate of the fractional abundance, which changes less significantly, especially for the values of correlation coefficient, $\gamma_{cr}$, in our case. This analysis tells us that the fractional abundance is not highly sensitive to other terms in the PDF expansion, in contrast to the primordial NGs, which determine the domain and the scalar power spectrum amplitude, which is present inside the variances and gives rise to an exponential sensitivity to the fractional abundance.

We now present some general discussions regarding our analysis based on the non-linearities, NGs, and the necessary perturbativity conditions,
\begin{itemize}[label=$\bullet$]
    \item \underline{\textbf{Inclusion of non-linearities}}: The broad analysis done in this letter does not ignore the apparent non-linearities in the density contrast field inherited from the possible non-linear behaviour of the curvature perturbations on super-horizon scales. We chose the compaction function as it provides a much better estimate for the formation threshold of the density contrast in the presence of the non-linear modifications \cite{Musco:2020jjb, Musco:2018rwt}.
    In this work, we have utilized this estimate to evaluate the PBH mass fraction using the critical scaling relation and the integration domain dependent on the threshold value and the impacts induced by non-linearities.  These non-linearities also do not tend to overpower their effects and become uncontrollable, which is visible in our previous results including the analysis of the figures \ref{ampvsfnl}-\ref{fpbhvamp}.
    \item
    \underline{\textbf{Effects of non-Gaussianities}}: These originate in our present analysis from two essential places. The first comes through the non-linearities in the density contrast field from where the non-Gaussian effects propagate, even when one assumes Gaussian statistics for the curvature perturbation. The second is connected with the generation of PBHs, where slow-roll violation for a brief period occurs to sufficiently enhance the scalar power spectrum amplitude.
    One must include these sources for a more realistic analysis of the PBH formation and their fractional abundance for the present-day total dark matter content. The local primordial NGs are present inside our use of the quadratic model for the curvature perturbation, which then gets translated into the compaction function. The amount of NG ($f_{\rm NL}$) present, hence, ultimately controls the scalar power spectrum amplitude and the PBH abundance, which will be elaborated on, shortly in the results of this section. 
    \item
    \underline{\textbf{Preserving perturbativity}}: In our setup, a USR phase to generate the large curvature perturbations comes with certain conditions imposed on its period. We have maintained throughout our analysis the critical condition on the e-foldings of the USR, $\Delta{\cal N}_{\rm USR}\sim {\cal O}(2)$ which guarantee that the cosmological perturbative approximations and controlled production of NGs are both satisfied. Another necessary perturbative approximation follows while restricting our analysis for the volume-averaged density contrast threshold to, $2/5 \leq \delta_{\rm th} \leq 2/3$, which is the same for the compaction threshold ${\cal C}_{\rm th}$ at the time of horizon crossing. In this window, non-linear effects present during the long-wavelength approximation on super-horizon scales remain controllable. The upcoming discussions on our results will show that preserving perturbativity brings about satisfactory outcomes.
    \item
    \underline{\textbf{Respecting Causality and Unitarity}}: The effective sound speed parameter, $c_{s}$, takes on a conformal time-dependent behaviour in our present setup. Its value at the pivot scale is given by $c_{s}(\tau_{*})=c_{s}$, where $\tau_{*}$ is the conformal time for the pivot scale $k_{*}=0.02 {\rm Mpc}^{-1}$. During our analysis of the SIGWs generated from Galileon theory, the properties of unitarity and causality have been preserved by satisfying the observational constraint, $0.024 \leq c_{s} < 1$, \cite{Planck:2015sxf}. 
     \item
    \underline{\textbf{Mass of PBH and NGs}}: PBH formation requires a brief period of slow-roll violation, which, in the present context, is introduced through a USR phase. Specifically, the wavenumber associated with the transition from the slow-roll to the USR phase, $k_{s}$, is the crucial parameter determining the resulting mass of the PBH. The dependency goes as $\tilde{c_{s}}k_{s} \propto 1/\sqrt{M_{\rm PBH}}$. Regardless of the exact position of the USR, large enhancements in the NGs remain present in our setup. In this work, we have set the transition scale at $k_{s}=10^{7}{\rm Mpc}^{-1}$, which corresponds to the formation of $M_{\rm PBH} \sim {\cal O}(0.01)M_{\odot}$, and additionally generates large negative NGs \cite{Choudhury:2023kdb}. 
\end{itemize}

\bibliography{references2}
\bibliographystyle{utphys}


\end{document}